\documentclass[article,twocolumn,prb,showpacs,superscriptaddress]{revtex4-2}

\usepackage{graphicx}
\usepackage{amssymb}
\usepackage{bm}
\usepackage{amsmath}
\usepackage{xfrac}
\usepackage{color}
\usepackage{titlesec}
\renewcommand{\figurename}{FIG.} 
\makeatletter	        
\def\fnum@figure{\textbf{\figurename~\thefigure}}
\makeatother
\usepackage[normalem]{ulem}
\usepackage[utf8]{inputenc}
\usepackage[T1]{fontenc}
\usepackage[marginal]{footmisc}

\usepackage{latexsym}

\usepackage[colorlinks=True,
                 linkcolor=black,
	        citecolor=blue,
	        urlcolor=blue]{hyperref}

\makeatletter	        
 \def\section{%
  \@startsection{section}{1}{\z@}{0.8cm plus1ex minus.2ex}{0.2cm}%
  {%
   \small\sffamily\bfseries\selectfont
   \raggedright
   \parindent\z@
  }%
 }%
  \def\subsection{%
  \@startsection{subsection}{2}{\z@}{0.8cm plus1ex minus.2ex}{0.2cm}%
  {%
   \small\sffamily\bfseries
   \raggedright
   \parindent\z@
  }%
 }%
\makeatother

\renewcommand\c{{\bf c}}

\newcommand{\comment}[1]{\textcolor{black}{#1}}

\usepackage{color}

\newcommand{\MOE}{MOE Key Laboratory for Nonequilibrium Synthesis and Modulation of Condensed Matter, Shaanxi Province Key Laboratory of Advanced Materials and Mesoscopic Physics, School of Physics, Xi’an Jiaotong University, Xi’an,710049, China}
\newcommand{\JAP}{Research Center for Electronic and Optical Materials, National Institute for Materials Science, Tsukuba 305-0044, Japan}
\newcommand{\JAPP}{Research Center for Materials Nanoarchitectonics, National Institute for Materials Science, Tsukuba 305-0044, Japan}

\makeatletter
\g@addto@macro\bfseries{\boldmath}
\makeatother

\begin{document}
\title{Topological Valley Transport in Bilayer Graphene Induced by Interlayer Sliding}
\author{Jie Pan}
\author{Huanhuan Wang}
\author{Lin Zou}
\author{Xiaoyu Wang}
\author{Lihao Zhang}
\author{Xueyan Dong}
\author{Haibo Xie}
\author{Yi Ding}
\author{Yuze Zhang}
\affiliation{\MOE}
\author{Takashi Taniguchi}
\affiliation{\JAPP}
\author{Kenji Watanabe}
\affiliation{\JAP}
\author{Shuxi Wang}
\email{shuxi.wang@xjtu.edu.cn}
\author{Zhe Wang}
\email{zhe.wang@xjtu.edu.cn}
\affiliation{\MOE}

\begin{abstract}
Interlayer sliding, together with twist angle, is a crucial parameter that defines the atomic registry and thus determines the properties of two-dimensional (2D) material \comment{homobilayers}. Here, we theoretically demonstrate that \comment{controlled} interlayer sliding in bilayer graphene induces Berry curvature reversals, leading to topological states \comment{confined} within a one-dimensional moir{\'e} channel. We experimentally realize interlayer sliding by bending \comment{the} bilayer graphene geometry across a nano-ridge. Systematic electronic transport measurements reveal topological valley transport when the Fermi energy \comment{resides} within the band gap, consistent with theoretical predictions of eight \comment{topological} channels. Our findings establish interlayer sliding as a powerful tool for tuning the electronic properties of bilayer graphene and underscore its potential for broad application across 2D material systems.



\end{abstract}

\maketitle

The layer degree of freedom is the unique physical freedom offered by the two-dimensional (2D) materials due to its weak interlayer van der Waals (vdW) bonding force. This \comment{distinctive characteristic} has facilitated the creation of diverse vdW heterostructures\cite{vdW1,vdw2,vdw3}, where stacking different types of 2D materials leads to modified properties via interface effects\cite{proximity2,proximity1}and generates new functionalities\cite{proximity3}. When applied to vdW homostructures—stacks of the same 2D material—this layer degree of freedom has revealed a range of novel physical phenomena. These include the Mott metal-insulator transition\cite{Cao2018mott}, unconventional superconductivity\cite{Cao2018sc}, 
 ferromagnetism\cite{sharpe2019}, topological phases\cite{Choi2021,Saito2021}, and the fractional quantum anomalous Hall effect\cite{FQAHE1,FQAHE2,FQAHE3,FQAHE4}, all prominently observed in moiré systems
such as magic-angle bilayer graphene and certain transition metal dichalcogenides (TMDCs). These findings underscore the potential of atomic registry manipulation via twist angle to dictate the physical properties of homostructures. However, beyond twist angles, interlayer sliding represents another critical parameter affecting the atomic registry of vdW systems that remains relatively unexplored.

\begin{figure}[t]
\includegraphics[width =1\linewidth]{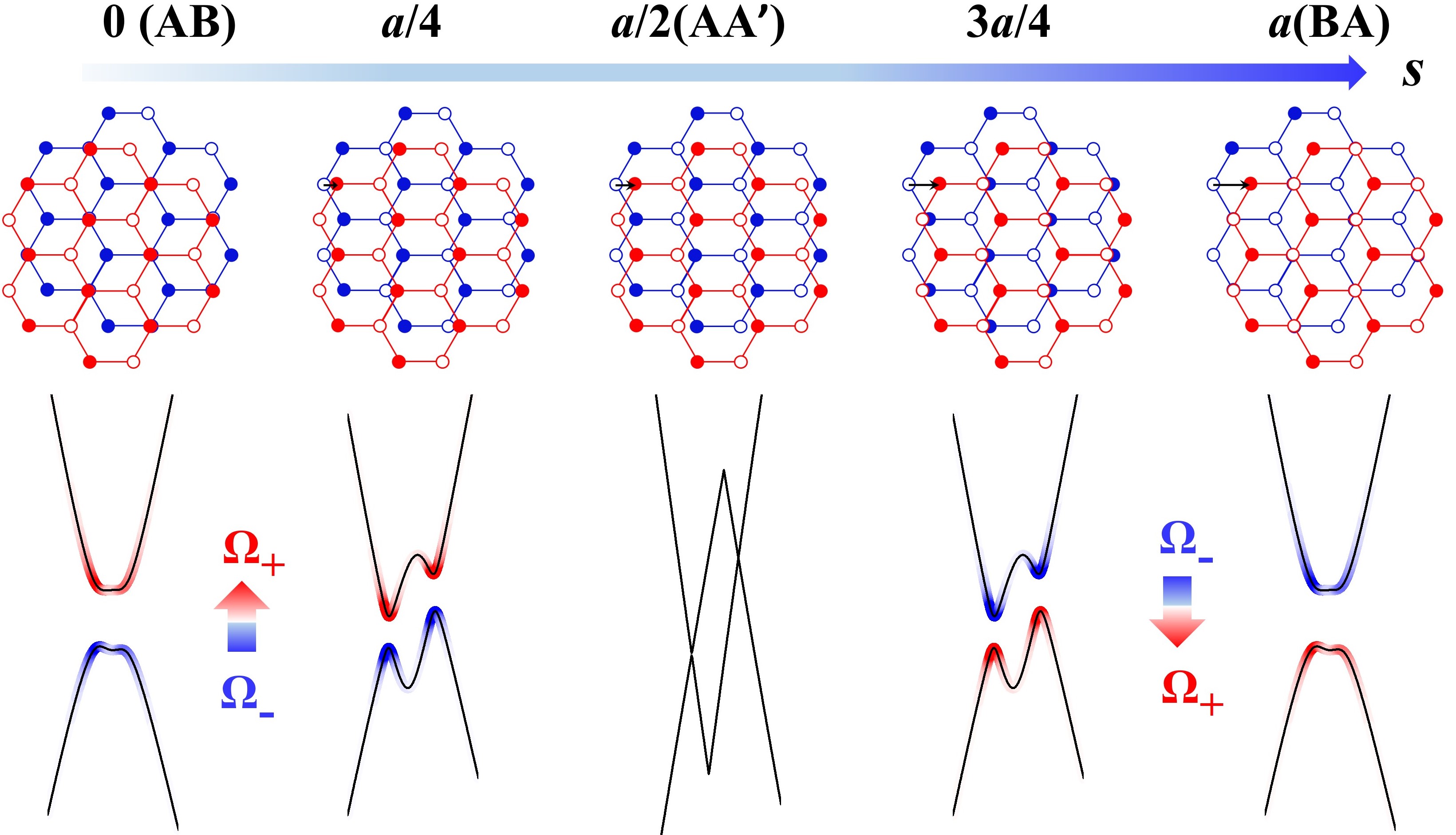}
\caption{\textbf{Bilayer graphene with interlayer sliding.}  Schematic illustration of bilayer graphene lattice structure with interlayer sliding along armchair direction, with sliding distance varying from 0 to $a$. To open a band gap, we assign a weak layer-dependent potential difference, $V = \pm 0.02t$, where $t$ is the nearest-neighbor intralayer hopping strength. Bandstructure of bilayer graphene with interlayer sliding is exhibited by black curves in the lower panel with color red(blue) denotes the positive(negative) Berry curvature.}
\label{carton-sliding}
\end{figure}

Uniform interlayer sliding typically results in non-stable configurations due to energy costs spread throughout the system, unlike twisted structures in which extra energy only cost in part of moiré patterns. Recent study has shown that stretching vdW heterostructures on polyimide substrates can overcome the potential barrier and induce interlayer sliding\cite{slide2024}. However, this approach \comment{has yet only worked for thick flakes} and faced challenges in adapting to low-temperature measurements. So far, interlayer sliding has mainly been applied to induce new type of ferroelectricity in non-centrosymmetric systems such as artificially stacked h-BN\cite{Stern2021,Pablo2021} and TMDCs\cite{ferro1,ferro2,ferro3,ferro4}, where sliding is driven by an external electric field. In contrast, for centrosymmetric 2D materials like bilayer graphene, where out-of-plane polarization is absent, intriguing possibilities \comment{have} only been predicted theoretically yet\cite{Park2015,Lee2015,Nam2021,PJ2024}. Interlayer sliding in bilayer graphene is expected to have a significant impact on its electronic and topological properties, including band gap reduction\cite{Park2015,Lee2015}, the emergence of a nonzero Berry curvature dipole \cite{PJ2024}, and particularly Berry curvature reversal. It is predicted to yield topological states in regions where the Chern number changes sign, but has yet to be realized experimentally.

\begin{figure*}[t]
\includegraphics[width =0.8\linewidth]{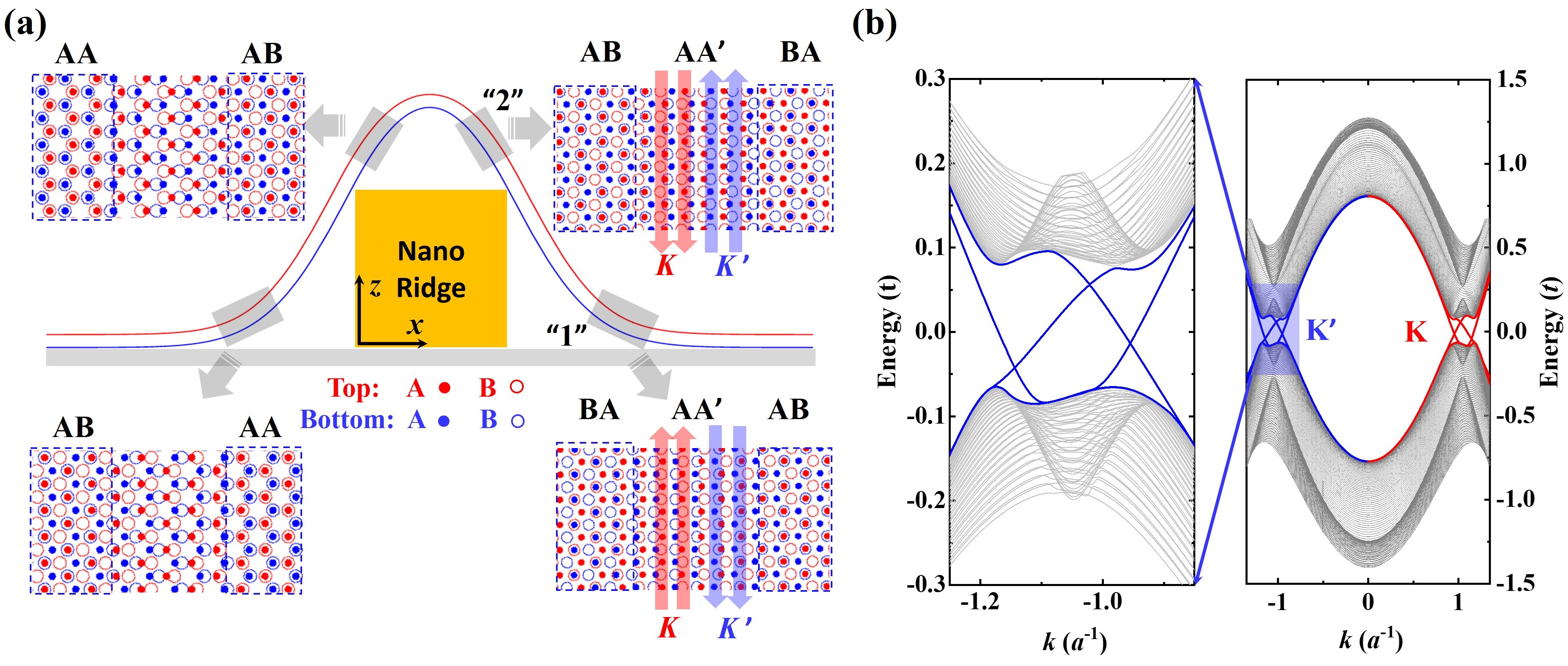}
\caption{\textbf{Bilayer graphene on nano-ridge.}  (a) Schematic illustration of experimental method for generating interlayer sliding in AB stacking bilayer graphene, forming AA$^\prime$ and BA stacking on right side, and AA/AB stacking on left side of the nano-ridge. (b) Bandstructure of bilayer graphene with interlayer sliding. Gray curves correspond to bulk states of AB or BA stacking and four red\comment{/blue} curves represent topological state at AA$^\prime$ moir{\'e} channel \comment{from K/K$^\prime$ valley}.}
\label{nanoridge}
\end{figure*}

In this study, we propose and experimentally implement a method to achieve interlayer sliding in bilayer graphene, that is, by placing it on top of the nano-ridge. We select the sliding direction along the armchair direction as shown in Fig. \ref{carton-sliding}, which is predicted to have the minimum energy barrier \cite{Park2015}. The interlayer sliding of this structure results in one-dimensional (1D) moir{\'e} channels, which host eight \comment{topological} states from two valleys. Ballistic transport behavior of corresponding topological valley states \comment{is} clearly observed in our low temperature electronic transport measurements. 



Fig.\ref{carton-sliding} illustrates the carbon lattice structure with varying interlayer sliding distances along the armchair direction and corresponding bandstructures, ranging from 0 to $a$ (where $a=0.142$ nm represents the carbon-carbon bond length). The magnitude of Berry curvature $\Omega$ is represented by color. To clearly demonstrate \comment{this}, a band gap is opened by assigning a weak layer-dependent potential difference. For a sliding distance $s<a/2$(close to AB stacking), the conduction band exhibits \comment{a} positive Berry curvature. AA$^\prime$ stacking ($s=a/2$) is the critical point, as \comment{the} system becomes gapless and \comment{the} Berry curvature is zero. Once the critical point is surpassed ($s > a/2$), the Berry curvature of \comment{the} conduction band reverses sign, becoming negative as the system approaches BA stacking, consistent with previous reports\cite{zhangfan2013,Ju2015}. Thus, a bilayer graphene system with continuously varying interlayer sliding experiences Berry curvature reversal, and the Chern number changes across the AA$^\prime$ stacking. Owing to \comment{the} bulk-edge correspondence, topological states are expected to emerge along sliding-induced AA$^\prime$ stacking.


When bilayer graphene is placed on top of a nano-ridge, as shown in Fig. \ref{nanoridge}(a), the geometry deformation would compel the bilayer graphene to slide. Actually, interlayer sliding has been experimentally observed in few-layer graphene under similar conditions, exhibiting superlubricity between adjacent layers\cite{Han2020}. The interlayer sliding $s$ can be described by the governing equation (as derived in the Supplementary Materials\footnote[100]{See Supplemental Material at [URL] for derivations of sliding distance, Monte Carlo simulations, and band structure simulations and additional experimental data, which include Refs \cite{youngs,AA2020,pseudo1,Ando}}),
\begin{equation}
    s = d_0 \arctan \left(-\frac{{\rm d}z}{{\rm d}x}\right).
\label{sliding1}
\end{equation}
Here $d_0 = 0.335$ nm is the interlayer separation\cite{Park2015}, $z$ and $x$ denote the vertical and horizontal \comment{coordinates}, respectively. \comment{Eq. \ref{sliding1} agrees quantitatively with our Monte Carlo simulations (see Supplementary Materials\footnotemark[\value{footnote}]) and accurately describes \comment{the} experimentally observed sliding distance in deformed few-layer graphene\cite{Han2020}.} By setting $s=a/2$, we get a critical slope $\left|\frac{{\rm d}z}{{\rm d}x}\right|$=0.21 \comment{required to} achieve AA$^\prime$ stacking.

Analysis of Eq. (\ref{sliding1}) reveals several general conclusions. The sliding distance $s(x)$ is an odd function as the bilayer graphene profile $z(x)$ is an even function of $x$, indicating that the stacking configuration differs on the left and right sides of the nano-ridge. To clearly demonstrate these, we specifically select maximum sliding $s = a$ and the stacking is BA on the right side while AA on the left side of the nano-ridge, as demonstrated in Fig. \ref{nanoridge}(a). However, it is important to note that AA stacking has the highest energy and is energetically unfavorable \cite{Park2015,reconstruction2}. Atomic reconstruction would likely occur to mitigate the AA stacking energy, similar to phenomena experimentally observed in twisted bilayer graphene systems\cite{reconstruction1}. \comment{To further verify this point, our Monte Carlo simulations demonstrate that the AB/AA$^\prime$/BA stacking configuration is energetically favorable in bent bilayer graphene, as it involves a significantly lower energy barrier for its formation and stability \cite{Park2015,reconstruction2,reconstruction1}, while the formation of AA stacking is strongly suppressed.} 

\begin{figure*}[t]
\includegraphics[width =0.9\linewidth]{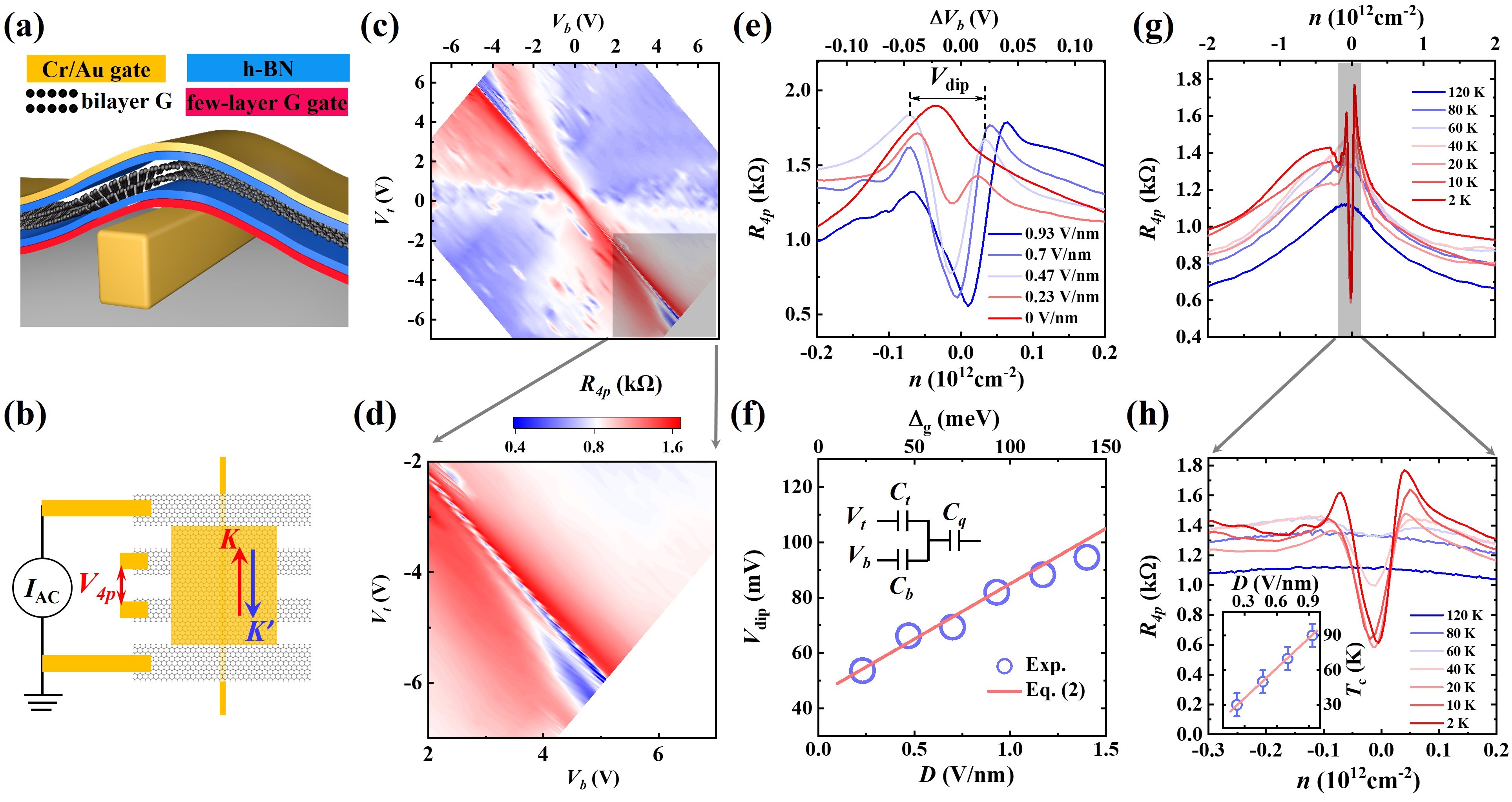}
\caption{\textbf{4-probe measurement results of bilayer graphene with interlayer sliding.}   Schematic illustrations of (a) experimental setup of heterostructures h-BN/bilayer graphene/h-BN/few-layer graphene and (b) 4-probe measurements.  At $T=2$ K, 4-probe resistance for different bottom and top gate voltages is summarized in (c) with enlarged view in (d). (e) For fixed $T=2$ K, 4-probe resistance as a function of carrier density for different displacement fields. (f) Measured $ V_{\rm dip}$ as a function of displacement field. 4-probe resistance as a function of carrier density for fixed displacement field $|D| =0.7$ V/nm under different temperatures are shown in (g) with enlarged view in (h). The inset demonstrates the positive correlation between critical temperature and displacement field.}
\label{4p}
\end{figure*}

Moreover, since the slope $\left|\frac{{\rm d}z}{{\rm d}x}\right|$ is 0 both at the center and far away from \comment{the} nano-ridge, the stacking configurations in these regions must be identical. As a result, AA$^\prime$ stacking, if exists, must appear in pairs in the system.  When the bilayer graphene deviates from the nano-ridge, \comment{a} 1D moir{\'e} channel is formed between \comment{the} AB and BA stacking sequence (right upper panel, region "2"), with AA$^\prime$ stacking in the center. Another pair of BA/1D moir{\'e}/AB with central AA$^\prime$ stacking (right lower panel, region "1") emerges as the bilayer graphene flattens out far from the nano-ridge. If an array of nano-ridges is present, periodical sliding can lead to 1D moir{\'e} pattern, similar to the theoretical proposal for lattice mismatched TMDCs heterostructures \cite{1dmoire}.

In our system, we expect four \comment{topological} states (without spin degeneracy), as each AB/BA stacking domain wall generates two chiral states per valley\cite{Martin2008,zhangfan2013,Ju2015,Helin2016,Li2016,zhu2024}. This analysis is corroborated by tight-binding calculations, with bandstructures shown in Fig. \ref{nanoridge}(b). The hopping parameterization follows Refs. \cite{Ando2001,Uryu2004,Trambly2010,Koshino2012}, as detailed in Supplementary Materials\footnotemark[\value{footnote}]. \comment{Within each domain wall, chiral states from opposite valleys counter-propagates; if one of the domain walls is suppressed (e.g., via intervalley scattering induced by atomic vacancies), the remaining one can still support chiral states and thereby producing valley-polarized currents.} Within either $K$ or $K^\prime$ valley, we identify four topological states, resulting from two spatially separated 1D moir{\'e} channels. Considering the spin degeneracy, eight topological channels along a single direction are expected to form along the interlayer sliding-induced AA$^\prime$ moir{\'e} channel. These \comment{topological} channels will lead to ballistic transport behavior when the Fermi level lies within the band gap, providing a signature of successful interlayer sliding in our system.

It is important to note that the expected number of topological channels in our system is twice that of \comment{the} naturally formed defected AB/BA domain walls\cite{Ju2015} and four times that of zigzag-terminated bilayer graphene edges\cite{zzedge,subgap2022}. This large number of topological channels can help eliminate these alternative mechanisms that also support ballistic transport. Additionally, when the sliding distance is less than $a$ (but larger than $a/2$ ), the resulting stacking will deviate from the ideal AB/AA$^\prime$/BA stacking configuration. Nevertheless, topological chiral states can still emerge under these conditions, as discussed in Ref. \cite{slidinggap} and Supplementary Materials\footnotemark[\value{footnote}].

Experimentally, we first fabricate Cr/Au nano-ridges with a width of 50 nm and a height of 65 nm on SiO$_2$/Si substrate by using e-beam lithography and e-beam evaporation. We assemble hBN/bilayer graphene/hBN/few-layer graphene with \comment{the} standard dry transfer technique, where \comment{the} few-layer graphene serves as bottom gate. When placing the heterostructures on the nano-ridge, we keep the nano-ridge parallel with the zigzag direction (such that the interlayer sliding is along \comment{the} armchair direction). Orientation of the bilayer graphene atomic carbon lattice is determined with Raman measurements\cite{Raman1} (see Supplementary Materials\footnotemark[\value{footnote}] for details). The sample is annealed in H$_2$ and Ar at 350 $^\circ$C for 3 hours to remove chemical residues and allow \comment{the} lattice to relax and induce interlayer sliding. Finally Cr/Au are deposited to create the top gate and contact electrodes, and the samples are etched by SF$_6$ and O$_2$ plasma to achieve a Hall bar geometry. The samples are loaded into a cryostat for low-temperature electronic transport measurements using standard AC techniques. An AC current (ranging from 1 to 10 nA) of frequency 17.377 Hz is applied and the AC voltage is measured by \comment{the} lock-in amplifier SR830.

We first conduct 4-probe measurements, as illustrated by Fig. \ref{4p}(b). In a dual-gate system, the carrier density is determined by $n=\varepsilon_r \varepsilon_0(V_b/d_b+ V_t/d_t)$ where $\varepsilon_r=3.5$ is the relative permittivity of h-BN\cite{hbn1,hbn2}, $V_{b(t)}$ denotes the applied gate voltages (with respect to the charge neutrality point (CNP) voltage) and $d_{b(t)}$ represents the thickness of the h-BN layers, which can be determined by measuring the Hall effect under varying gate voltages. The displacement field is given by $D =\varepsilon_r (V_t/d_t - V_b/d_b)$. The absolute amplitude of \comment{the} displacement field $|D|$ determines the band gap $\Delta_g$, with the relation $\Delta_g \approx 100 |D|$ for Bernal stacking bilayer graphene, where $\Delta_g$ is in the unit of meV and $|D|$ is in the unit of V/nm \cite{Zhang2009,Zhu2017,subgap2022}. 

The 4-probe resistance as a function of the bottom and top gates measured at 2 K is shown in Fig. \ref{4p}(c). The CNP forms a straight line defined by $n(V_b, V_t) = 0$, where high resistance is expected for Bernal stacking bilayer graphene \cite{Zhang2009}. However, our low temperature 4-probe measurements reveal a contrasting behavior. Instead of a high resistance near the CNP line,  we observe a significant resistance drop\cite{Campos2019}. This is more clearly demonstrated in \comment{the} enlarged view shown in Fig. \ref{4p}(d). This resistance drop is a strong signature for the ballistic transport of topological chiral states within the band gap, which enables current to flow with minimal scattering and leads to the sharp reduction of 4-probe resistance.

The resistance drop behavior is influenced by the displacement field (band gap size) of bilayer graphene. Fig. \ref{4p}(e) summarized the carrier density dependence of resistance for different displacement fields at 2 K. For large displacement field cases, the resistance drops at the CNP, which can be attributed to the suppression of scattering between topological and normal states as the bulk becomes insulating. When the displacement field decreases to zero, the resistance is maximum at the CNP without resistance drop. It is noted that as the displacement field increases, the range $V_{\rm dip}$ ---over which the resistance drop occurs --- broadens. We summarize the experimental results for $V_{\rm dip}$ in Fig. \ref{4p}(f). 

The gate voltage range $V_{\rm dip}$ corresponds to \comment{the} Fermi level variation $\Delta_{\rm eff}$ in which the transport is ballistic. To evaluate this variation in a gaped system, we need to consider the quantum capacitance $C_q = n_0 e^2 / \Delta_{\rm eff}$, where $n_0$ denotes the density of defect states within the gap\cite{Meric2008}. Using the circuit structure of capacitors shown in the inset of Fig. \ref{4p}(f), we establish a connection between gate voltage range $V_{\rm dip}$ and corresponding Fermi level variation $\Delta_{\rm eff}$,
\begin{equation}
    eV_{\rm dip} =\frac{C_b+C_t}{2C_b} \Delta_{\rm eff}+\frac{n_0e^2}{2C_b},
\label{cap}
\end{equation}
where $C_{b(t)}$ represents geometric bottom(top) capacitance. Our experimental observation of \comment{the} linear relation between $V_{\rm dip}$ and $D$ (shown in Fig. \ref{4p}(f)) \comment{implies} that $\Delta_{\rm eff}= \alpha D$, that is, the Fermi energy range of ballistic transport is proportional to the band gap size of bilayer graphene.  

By substituting $\Delta_{\rm eff}= \alpha D$ into Eq. (\ref{cap}), the experimental data \comment{are} well fitted as shown by the red line in Fig. \ref{4p}(f). The fitting yields $\alpha = 44$, and the finding that $\alpha < 100$ suggests $\Delta_{\rm eff}$ is narrower than $\Delta_{g}$, the band gap of the Bernal-stacked bilayer graphene. This discrepancy may arise from disorder-related smearing effects or reduction of the band gap in bilayer graphene under interlayer sliding\cite{PJ2024}. \comment{Additionally, while deformations of bilayer graphene can induce pseudomagnetic fields—which might also contribute to an overall band gap reduction—such fields cannot account for the topological channels observed under a significant displacement field \footnotemark[\value{footnote}].}


\begin{figure}[t]
\includegraphics[width =1\linewidth]{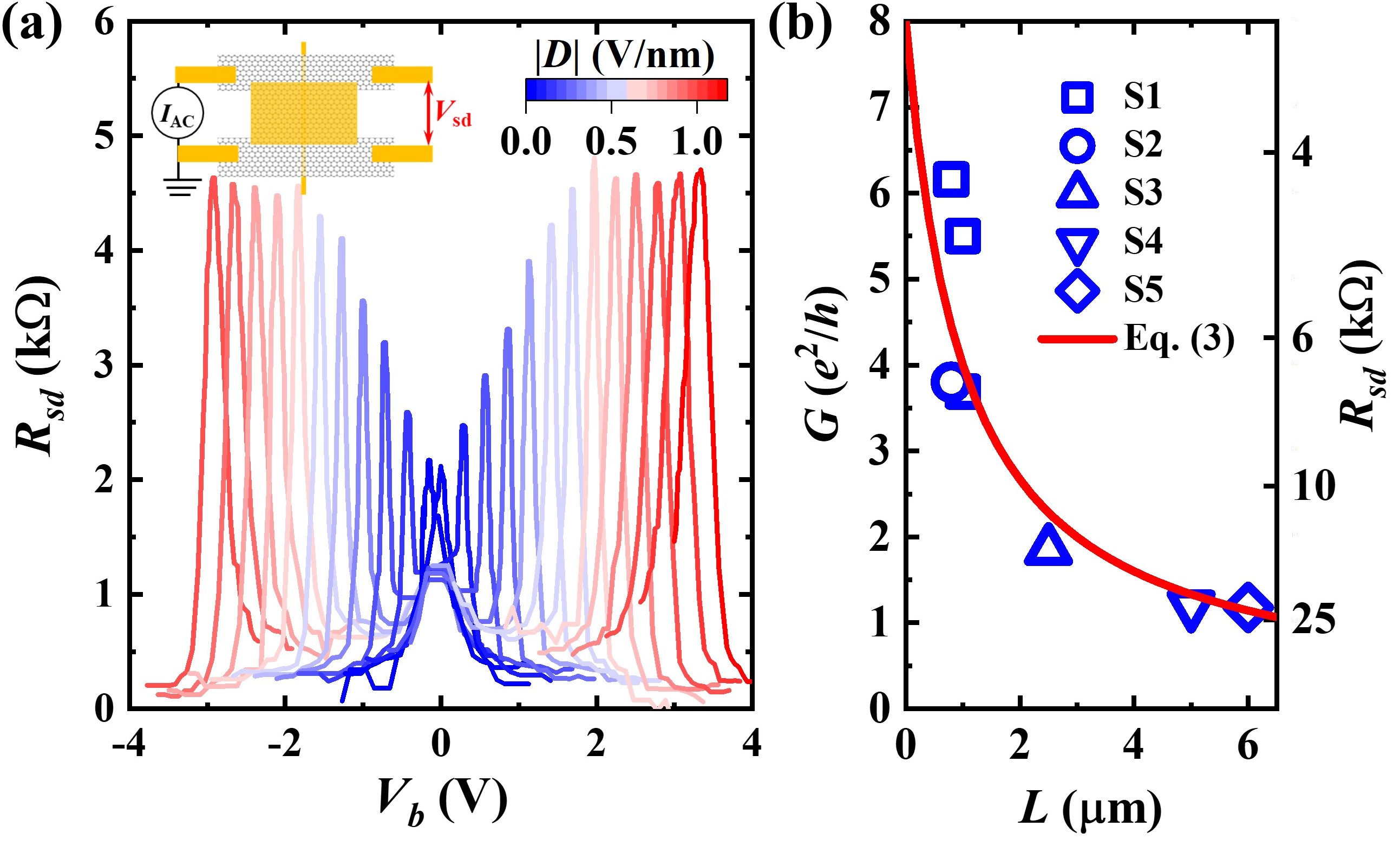}
\caption{\textbf{Source-drain resistance of bilayer graphene with interlayer sliding.}  (a) Source-drain resistance as a function of back gate voltage for different displacement fields. (b)  Open squares denote saturated conductance for different channel length $L$ of five samples and red curve represents fittings of Landauer-Büttiker formula with $L_{\rm mfp} = 1$ ${\rm \mu}$m.}
\label{2p}
\end{figure}

We further investigate the resistance drop behavior at different temperatures and Fig. \ref{4p}(g) shows the case with fixed $D = 0.7$ V/nm. At low temperatures, the resistance drop appears in a narrow carrier density region within $10^{11}$cm$^{-2}$ near the CNP, as shown in the enlarged view in Fig. \ref{4p}(h). The minimum resistance levels off at around 600 $\Omega$. The resistance drop becomes less profound above 60 K and totally \comment{disappears} above 80 K. Instead, resistance peaks near the CNP, similar \comment{to the} Bernal stacking bilayer graphene, are observed at higher temperatures (> 80 K). It is noted that for $D = 0.7$ V/nm, the effective gap $\Delta_{\rm eff} = 30$ meV is approximately four times the thermal energy of the critical temperature around 80 K. In addition, we observe a positive correlation between the critical temperature and the strength of the displacement field, as shown in the inset of Fig. \ref{4p}(h).

After demonstrating the ballistic transport by standard 4-probe measurements of Hall bar geometry, we switch to source-drain resistance ($R_{sd}$) measurements (inset of Fig. \ref{2p}(a)) to further investigate the topological channel numbers. Fig. \ref{2p}(a) \comment{shows} the back gate voltage dependence of the source-drain resistance for different displacement fields. For each curve, the top gate voltage is varied in conjunction with the back gate voltage to maintain a constant displacement field strength. The CNP resistance increases as the displacement field $|D|$ increases from 0 to 0.6 V/nm, and \comment{saturates} around 4.7 k$\Omega$ when $|D| > 0.6$ V/nm.  4.7 k$\Omega$ is much lower than the CNP resistance typically measured for gapped bilayer graphene \cite{Zhang2009,Pablo2010,Ju2015,Zhu2017,subgap2022}, close to the expected resistance $h/8e^2 = 3.2$ k$\Omega$ of 8 \comment{topological} channels. This saturation behavior is again consistent with the emergence of topological transport channels within the band gap of bilayer graphene \cite{Ju2015,Zhu2017,Campos2019}.

The saturation resistance of 4.7 k$\Omega$ being larger than expected $h/8e^2 = 3.2$ k$\Omega$ suggests the presence of finite intervalley scattering between chiral transport within one domain wall moving in opposite directions\cite{Ju2015}. In this case, the saturation resistance is expected to depend on the channel length. Five samples show consistent topological valley transport and their saturated conductance as a function of the channel length $L$ are summarized in Fig. \ref{2p}(b). A clear increase in conductance is observed as the channel length decreases. To model the channel length dependence, we apply the 1D Landauer-Büttiker formula\cite{Ju2015,Li2016,zhu2024,Campos2019,subgap2022},
\begin{equation}
    G = \frac{8e^2/h}{1+L/L_{\rm mfp}},
\label{LB}
\end{equation}
where $L_{\rm mfp}$ is the mean free path that characterizes the scattering between 1D \comment{topological} transport channels. \comment{Following Refs\cite{Ju2015,Li2016,Campos2019}, we neglect the contact resistance (which originates from straying current effects of regions without top gates\footnotemark[\value{footnote}]).} Eq. (\ref{LB}) fits well our experimental data, as represented by the solid red curve in Fig. \ref{2p}(b); thus it strongly supports the conclusion of 8 \comment{topological} channels in our system. \comment{This fitting yields $L_{\rm mfp}=$1 $\mu$m, consistent with values reported for hBN encapsulated bilayer graphene systems.\cite{Li2016,Chen2020}}


In summary, we successfully induce interlayer sliding in the bilayer graphene by using the nano-ridge, which leads to the formation of 1D moir{\'e} channels thus host eight topological channels. These are demonstrated by the 4-probe resistance drop and source-drain resistance saturation when \comment{the} Fermi level is tuned into the gap of our system. \comment{Pairs of domain walls provide new platform for exploring novel phenomena, for example, current injection across the nano-ridge could produce conductance oscillations under small magnetic fields, analogous to the new version of Aharonov-Bohm effect observed in monolayer graphene p-n junctions with quantum Hall edge states\cite{AB2, AB}.}  Our work underscores the potential of inducing interlayer sliding as \comment{a} new method to tune the atomic structures and electronic properties of the 2D material family. 

$Acknowledgement-$This work is financially supported by National Natural Science Foundation of China (Grants no. 12304232 and 12374121), Shaanxi Fundamental Science Research Project for Mathematics and Physics (Grant no. 23JSQ011),  China Postdoctoral Science Foundation No. 2023TQ0266, the Fundamental Research Funds for the Central Universities. K.W. and T.T. acknowledge support from the JSPS KAKENHI (Grants no. 21H05233 and 23H02052) and World Premier International Research Center Initiative (WPI), MEXT, Japan. We thank Yajun Zhang at Instrument Analysis Center of Xi'an Jiaotong University for the technical support.

\bibliography{biblio.bib}

\end{document}


\title{Supplementary Materials for ``\textit{Topological Valley Transport in Bilayer Graphene Induced by Interlayer Sliding}"}


\author{Jie Pan}
\author{Huanhuan Wang}
\author{Lin Zou}
\author{Xiaoyu Wang}
\author{Lihao Zhang}
\author{Xueyan Dong}
\author{Haibo Xie}
\author{Yi Ding}
\author{Yuze Zhang}
\affiliation{\MOE}
\author{Takashi Taniguchi}
\affiliation{\JAPP}
\author{Kenji Watanabe}
\affiliation{\JAP}
\author{Shuxi Wang}
\author{Zhe Wang}
\email{zhe.wang@xjtu.edu.cn}
\affiliation{\MOE}

\maketitle

\renewcommand\thesection{\arabic{section}}
\titleformat{\section}[block]
  {\normalfont\large\bfseries}
  {Supplementary Note \thesection:}{0.5em}{\centering}
\makeatother

\newcommand\4{$_4$}
\renewcommand\a{{\bf a}}
\renewcommand\b{{\bf b}}
\renewcommand\c{{\bf c}}
\renewcommand\d{{\bf d}}
\renewcommand{\theequation}{S\arabic{equation}}

\titleformat{\section}[block]
  {\normalfont\large\bfseries}
  {Supplementary Note \thesection:}{0.5em}{\centering}

\section{Interlayer sliding in deformed bilayer graphene} 
\subsection{Theoretical derivations}
We begin by deriving the interlayer sliding distance for bilayer graphene subjected to an arbitrary deformation. \comment{In the following derivation, we ignore possible carbon bond deformations, which proves to be a reasonable approximation when sliding across the AA$^\prime$ stacking configuration, as validated by subsequent Monte Carlo simulations and experimental observations.} To evaluate this sliding distance, we consider two infinitesimal arcs that represent the deformed bottom and top graphene layers, as illustrated in Supplementary Fig. \ref{arc}. We assume that T$_{1(2)}$ is exactly on top of B$_{1(2)}$ when the bilayer system is flat as shown in   Supplementary Fig. \ref{arc}(a). When bilayer system is deformed as shown in  Supplementary Fig. \ref{arc}(b), we can see that instead of T$_{2}$, T$_{3}$ is on top of B$_{2}$. The length between T$_{2}$ and T$_{3}$ characterizes the interlayer sliding distance $ds$.

\begin{figure}[H]
\centering
\includegraphics[width =0.8\linewidth]{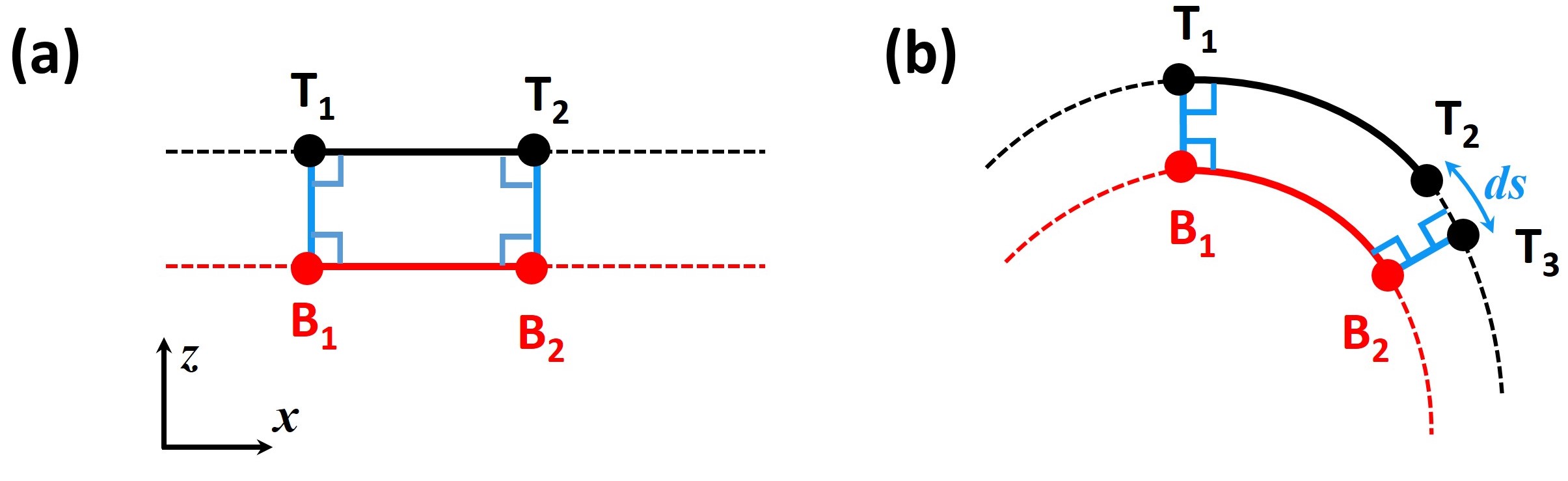}
\caption{\textbf{Illustration of interlayer sliding due to bending.} (a) Flat bilayer system, where T$_{1(2)}$ is exactly on top of B$_{1(2)}$. (b) Deformed bilayer system, where T$_{1}$ is on top of B$_{1}$ while T$_3$ is on top of B$_2$.}
\label{arc}
\end{figure}

We assume that the profile of the bottom graphene layer is described by the function $z(x)$. The coordinates of site B$_1$ can then be expressed as $(x, z(x))$ and site B$_2$ as $(x+dx, z(x+dz))$. The length of the arc $\arc{B_1B_2}$ can then be evaluated as:
\begin{equation}
    l_{\arc{B_1B_2}}=\sqrt{1+z'(x)^2}dx.
    \label{b1b2}
\end{equation}

To evaluate the length of $\arc{T_1T_3}$, we first obtain the coordinates of site T$_1$ and T$_3$. It should be noted that the vector $B_1T_1$ and $B_2T_3$  is perpendicular to the arcs $\arc{B_1B_2}$ and $\arc{T_1T_3}$. Therefore the coordinates of sites T$_1$ and T$_3$ can be determined as follows,
\begin{equation}
\begin{cases}
{\rm T}_1 = \left(x-\frac{d_0z'(x)}{\sqrt{1+z'(x)^2}},z'(x)+\frac{d_0}{\sqrt{1+z'(x)^2}} \right) \\
{\rm T}_3 = \left(x+dx-\frac{d_0z'(x+dx)}{\sqrt{1+z'(x)^2}},z(x+dx)+\frac{d_0}{\sqrt{1+z'(x)^2}} \right) \\
\end{cases},
\end{equation}
where $d_0$ is the interlayer separation, which is treated as a constant in our case. Using the coordinates of T$_1$ and T$_3$, we can now derive the length of $\arc{T_1T_3}$,
\begin{equation}
l_{\arc{T_1T_3}}=\left(1-\frac{d_0z''(x)}{(1+z'(x)^2)^{3/2}}\right)\sqrt{1+z'(x)^2}dx.
\label{t1t2}
\end{equation}

Subtracting Eq. \ref{b1b2} from Eq. \ref{t1t2}, we can derive the sliding distance for this infinitesimally arc,
\begin{equation}
ds=-\frac{d_0z''(x)}{\sqrt{1+z'(x)^2}}dx=u(x)dx.
\label{ds}
\end{equation}
After integrating out Eq.\ref{ds}, we obtain the total interlayer sliding distance,
\begin{equation}
s=d_0 \arctan(-z'(x)),
\label{st}
\end{equation}
which is exactly Eq. (1) in the main text. Based on Eq. \ref{st}, we conclude that the accumulated sliding distance at a given position is determined by the local slope $z'(x)$.

\subsection{Monte Carlo simulations}

\comment{In the following, we use Monte Carlo simulations to demonstrate that, Eq. \ref{st} can accurately describe the interlayer sliding distance when the sliding energy barrier is low (for AB/AA$^\prime$/BA crossover) but fails when the energy barrier is high (for AB/AA crossover), as illustrated in Supplementary Fig. \ref{potential}. When bilayer graphene is deformed, the system can respond via in-plane strain (with strain field $\epsilon(x)$) or via interlayer sliding (described by a local sliding field $\delta(x)$). These fields obey the geometric constraint,}

\begin{equation}
u(x)=\epsilon(x)+\delta (x).
\label{deform}
\end{equation}

\comment{The presence of a nonzero strain field $\epsilon(x)$ contributes to the elastic energy, given by $E_\epsilon=\int \frac{1}{2}E_{2D}W\epsilon^2dx$, where $E_{2D}$ is 2D Young’s modulus of graphene, taken as $340$ N/m\cite{youngs}, $W$ represents the width of the graphene sheet. Meanwhile, interlayer sliding modifies the local stacking configuration, leading to a change in stacking energy. This energy contribution is expressed as $E_s=\int E_{0}(s)/A_{\rm unit}W dx$, where $A_{\rm unit}$ is the area of the unit cell and $E_0$ is the stacking energy per unit cell \cite{Park2015} as shown in Supplementary Fig. \ref{potential}. $E_0$ is dependent on the sliding distance $s$, which is the accumulated sliding distance and related to the sliding field $\delta (x)$ by $s=\int \delta (x) dx$.}

\begin{figure}[H]
\centering
\includegraphics[width =0.5\linewidth]{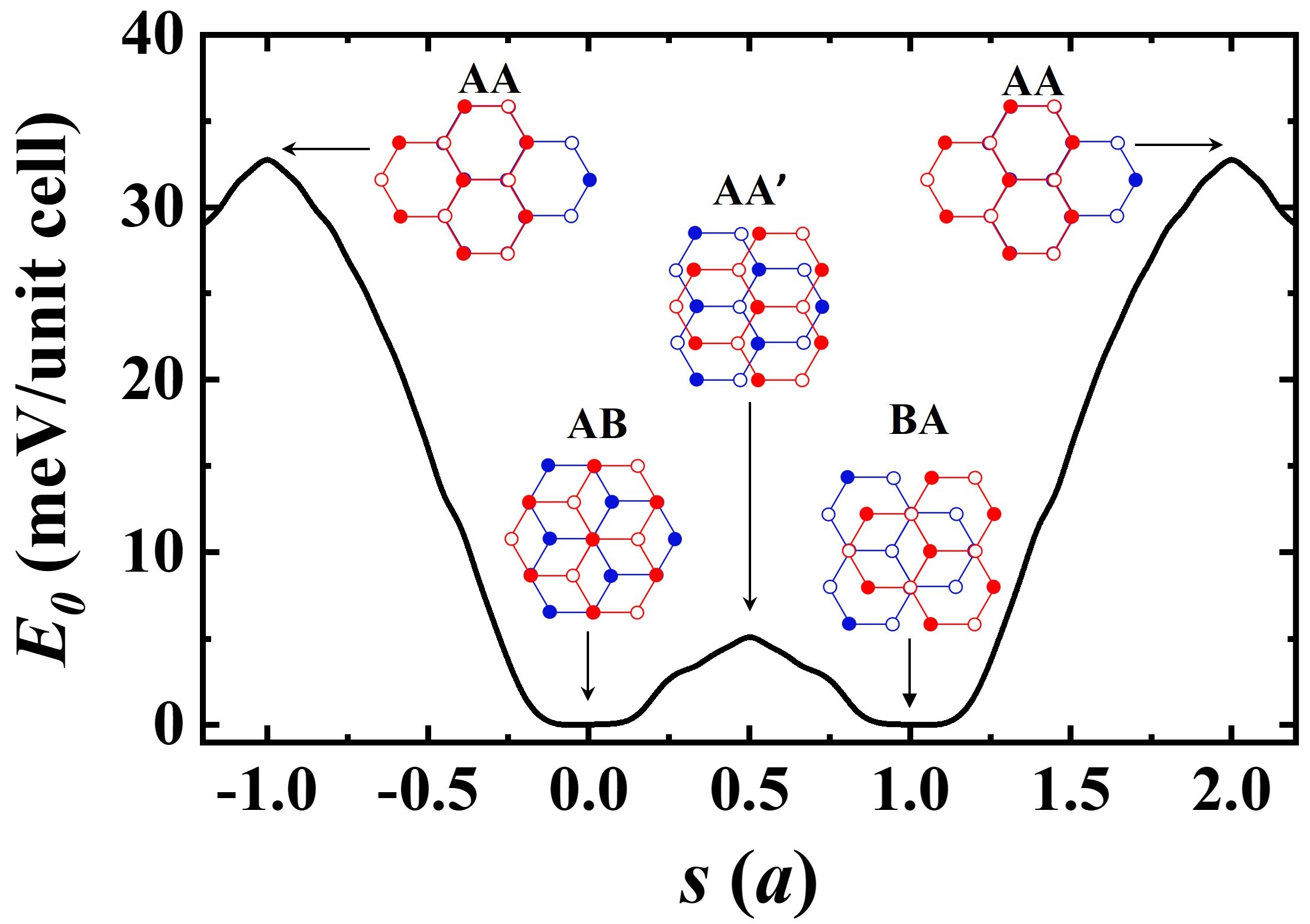}
\caption{\textbf{Stacking energy per unit cell under different sliding distances.} The stacking energy at different stacking configurations are adapted from Ref. [\onlinecite{Park2015}].}
\label{potential}
\end{figure}

\comment{In the Monte Carlo simulation, we assume the deformed bilayer graphene sheet to be in the shape of $z(x)=h_0\exp(-\frac{x^2}{2\lambda^2})$, with $h_0=50$ nm, $\lambda=$ 60 nm. The annealing algorithm is applied and total enery $E_t=E_\epsilon+E_s$ is minimized. Starting from various initial conditions, the system consistently converged to the same minimum energy configuration after $2\times10^9$ steps as shown in Supplementary Fig. \ref{monte}(a). The resulting sliding field $\delta(x)$ is presented as the black curve in the upper panel of Supplementary Fig. \ref{monte}(b), while the red curve represents the analytical solution obtained by $\delta(x)=u(x)=-\frac{d_0z''(x)}{\sqrt{1+z'(x)^2}}$. In the lower panel of Supplementary Fig. \ref{monte}(b), we plot the accumulated sliding distance $s(x)$ as a function of position. On the $x>0$ side, we find excellent agreement between the simulation (black curve) and analytical results (red curve) given by $s(x)=d_0 \arctan(-z'(x))$. In this case, we can observe that there are two AA$^\prime$ stacking regions. In contrast, on the $x<0$ side, the interlayer sliding distance $s$ based on Monte Carlo simulations (black curve) is almost zero. This strong suppression of sliding is attributed to the high-energy barrier associated with AA stacking. We conclude that the analytical solution (Eq. (1) in the main text) accurately describes the sliding behavior on the $x>0$ side, where the relevant energy barriers are comparatively low.}

\begin{figure}[H]
\centering
\includegraphics[width =0.8\linewidth]{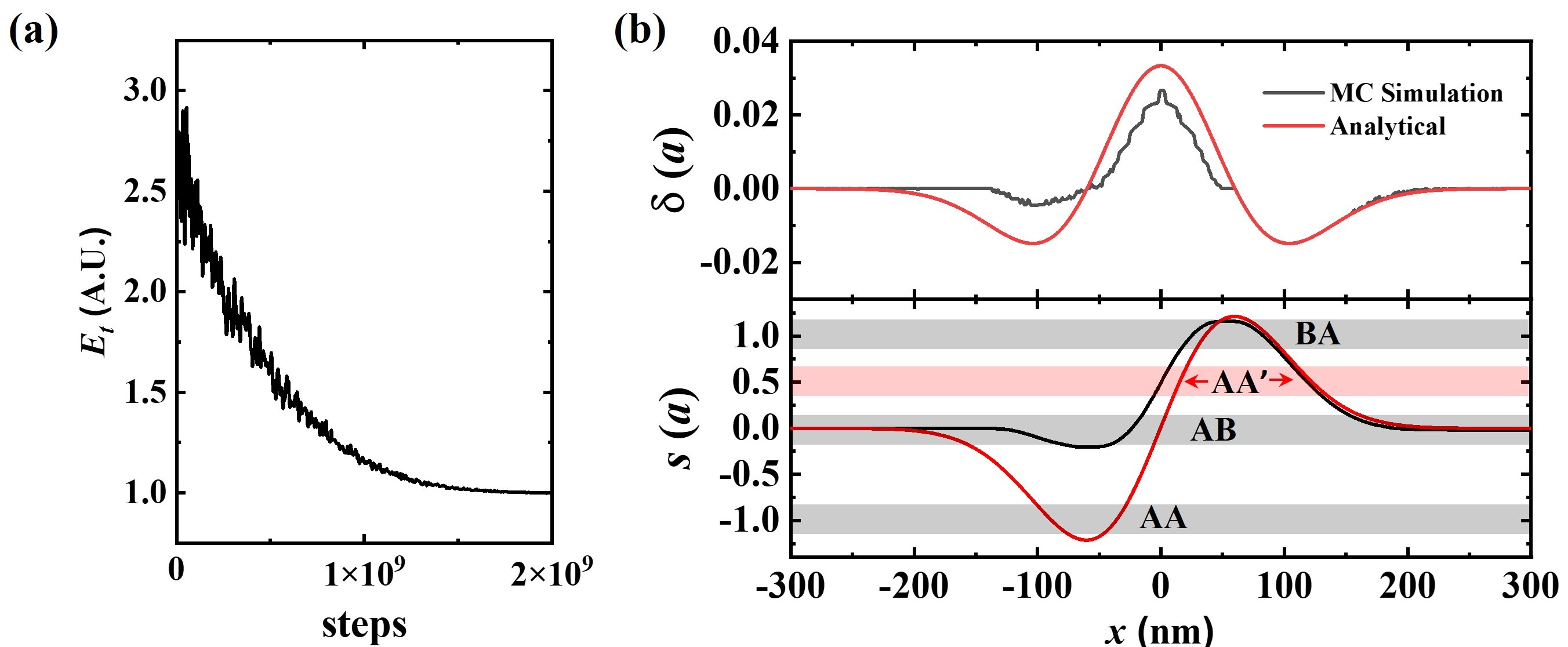}
\caption{\textbf{Monte Carlo simulations of interlayer sliding.} (a) The total energy $E_t$ as a function of iteration steps. (b) Simulation results for sliding field $\delta(x)$ and accumulated sliding distance $s(x)$.}
\label{monte}
\end{figure}

\comment{Therefore, the Monte Carlo simulation confirms that Eq. \ref{st} can capture the interlayer sliding on the $x>0$ side and predict the formation of a pair of AB/AA$^\prime$/BA structures. On the $x<0$ side, the AA stacking has a much higher energy barrier, and consequently, the interlayer sliding is largely suppressed. It should be noted that interlayer sliding in deformed few-layer graphene has been directly observed by TEM\cite{Han2020}. A sliding distance of $a\sim2a$ is observed for tetralayer graphene deformed by $12\degree$. Based on Eq. \ref{st} with $d_0=7.2a$, we obtain the sliding distance $1.5a$, which aligns well with the experimental observations. Therefore, the theoretical description of the sliding distance is quite reasonable, as confirmed by both Monte Carlo simulations and comparison with experimental observations.}

\clearpage
\newpage

\section{Tight-binding simulations of bandstructure of deformed bilayer graphene}
\subsection{Bilayer graphene with interlayer sliding}
\begin{figure}[H]
\centering
\includegraphics[width =0.7\linewidth]{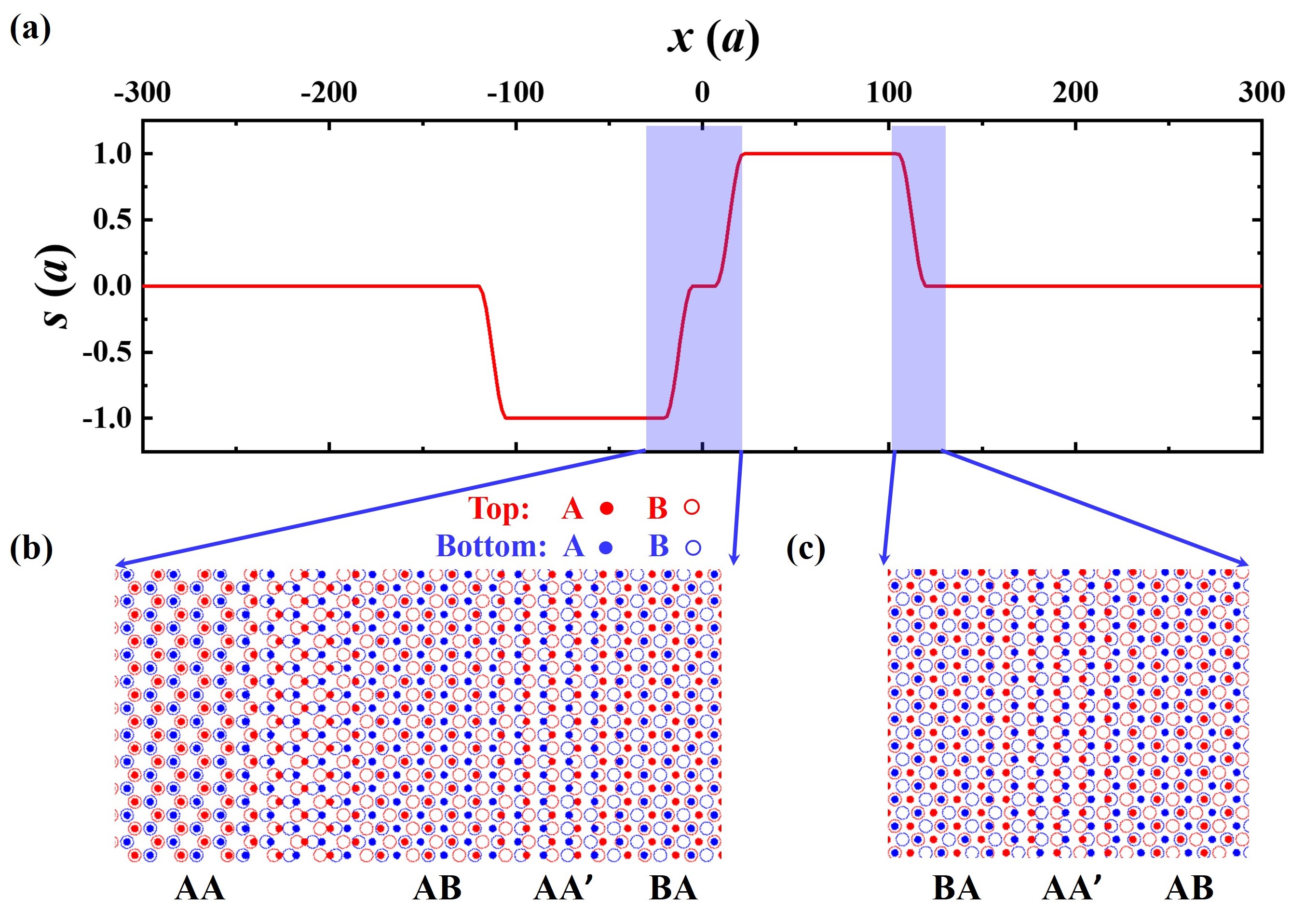}
\caption{\textbf{(a) Interlayer sliding distance profile and (b) stacking configurations.}}
\label{sliding}
\end{figure}

Based on the derivations in Supplementary Note 1, we assign the interlayer sliding distance for bilayer graphene on top of a nano-ridge. The sliding distance $s(x)$ is an odd function of $x$, corresponding to the even symmetry of $z(x)$. This symmetry difference leads to distinct stacking configurations on different sides of the nano-ridge. We modeled sliding along the armchair direction, setting the maximum sliding distance to $a$, which forms an ideal AB/BA stacking sequence on the right side of the nano-ridge (Supplementary Fig. \ref{sliding}). Notably, the AB-to-BA transition on the right has a low energy barrier \cite{Park2015}. However, on the left side, AB-to-AA transition is unlikely to occur in real devices, as it is energetically unfavorable\cite{Park2015,reconstruction1}. Thus, we do not expect the AA stacking in our bilayer graphene with interlayer sliding, as discussed in the main text.

The analysis of Eq. \ref{st} reveals that AA$^\prime$ moir{\'e} channels must appear in pairs. Since $z'(x)=0$ when $x=0$ and as $x$ approaches regions far from the nano-ridge, the stacking reverts to the same AB configuration. This condition leads to the formation of paired AA$^\prime$ channels, as shown in Supplementary Fig. \ref{sliding}(b\&c).

After assigning the sliding distance profile, we can construct the Hamiltonian using a tight-binding model,
\begin{equation}
H = \sum_{\{i,j\}} t\left(\bm {r_{i}-r_{j}} \right)\left(c_i^\dagger c_j +c.c.\right) + \sum_{i} V_\pm c_i^\dagger c_i .
\label{ham}
\end{equation}
where $c_i^\dagger$ and $c_j$ denote creation and annihilation operators, $V_\pm$ represents the layer potential difference. The hopping strength in Eq. \ref{ham} is given by\cite{Ando2001,Uryu2004,Trambly2010,Koshino2012},
\begin{equation}
t\left( {\boldsymbol{\delta },z} \right) = {V_\pi } e^  { - \left( {d - a} \right)/{r_0}} {\sin ^2}\alpha  + {V_\sigma } e^ { - \left( {d - d_0} \right)/{r_0}} {\cos ^2}\alpha.
\label{hopping}
\end{equation}
where $V_\pi = - t$ is the intralayer hopping component from the $\pi$ bond, and $V_\sigma = 0.18t$ is the interlayer hopping component from the $\sigma$ bond. $d$ is the distance between two atoms, which is defined as $d =\sqrt{\left| \boldsymbol{\delta} \right|^2+z^2}$. Here $|\boldsymbol{\delta}|$ is the intralayer distance and $z$ is the interlayer displacement component respectively, and the constant $r_0 = 0.0453$ nm is the decaying length. At last, $\alpha$ is the angle between the displacement vector $(\boldsymbol{\delta}, z)$ and the $\boldsymbol{z}$ axis. 

By substituting Eq. \ref{hopping} into Eq. \ref{ham}, we can numerically obtain the Hamiltonian, and the corresponding bandstructures are derived by diagonalizing the Hamiltonian matrix. In our simulation, we set $V=0.2t$ and choose an intralayer cutoff distance $\delta=1.5a$. To eliminate states arising from the edges of the bilayer graphene ribbon, we applied periodic boundary conditions to connect the left and right sides of the ribbon. In the following, we simulate the AB/AA$^\prime$/BA stacking and AA/AB stacking configurations separately. The simulation results are summarized in Supplementary Fig. \ref{bs}. We present the bandstructures near K' valley. 

In panel (a), we consider the AB/AA$^\prime$/BA stacking pairs, where interlayer sliding occurs solely on the right side of the nano-ridge. Two pairs of chiral states clearly appear within the band gap, which corresponds precisely to what is illustrated in Figure 1(b) in the main text.

\begin{figure}[H]
\centering
\includegraphics[width =0.8\linewidth]{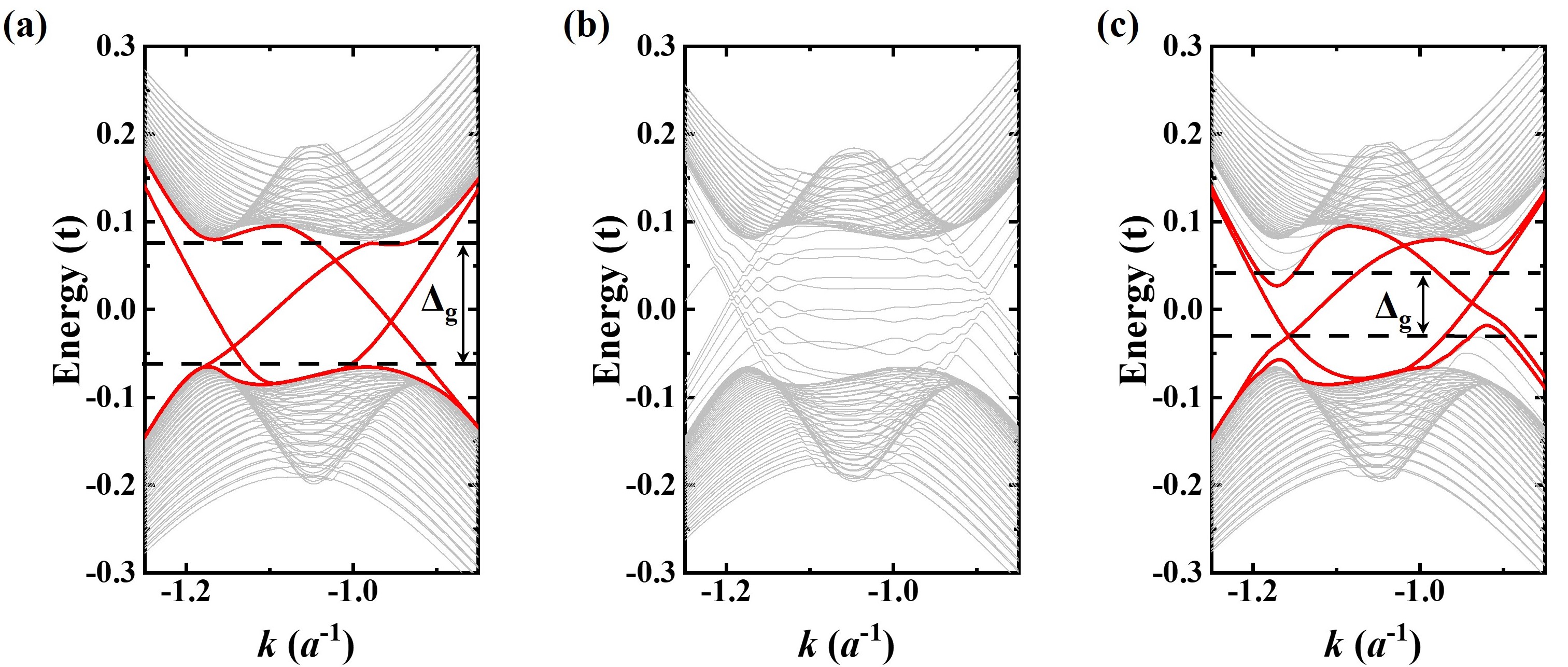}
\caption{\textbf{Simulated bandstructures of bilayer graphene with (a) AB/AA$^\prime$/BA, (b)AA/AB stacking pairs and (c) sliding distance $s=0.7a$}}
\label{bs}
\end{figure}

In panel (b), we focus on the AA/AB stacking domain wall pairs, where interlayer sliding is restricted to the left side of the nano-ridge. As noted in the main text and Monte Carlo simulations, AA stacking has the highest energy and is thus energetically unfavorable, meaning that AA stacking is unlikely to appear on the left of the nano-ridge in actual samples. Nevertheless, we can still simulate the bandstructures and observe a significant presence of midgap states, arising from the gapless AA stacking region\cite{AA2020}. 

In panel (c), we explore the bandstructure with an maximum interlayer sliding distance of $s = 0.7a$, where AA$^\prime$ stacking is formed as $s > 0.5a$, but the BA stacking configuration is not reached. Despite this, we still observe the emergence of topological bands within the band gap. It is clear that the band gap is smaller than that of the Bernal stacking configuration shown in panel (a). This reduction in the band gap may explain why the experimentally measured energy window that supports ballistic transport is smaller than band gap of AB-stacked bilayer graphene, as discussed in the main text.

\subsection{Bilayer graphene with pseudomagnetic field}
\comment{In the deformed graphene, the pseudo-vector potential is directly related to the strain tensor, which can arise from both in-plane and out-of-plane lattice displacements. As thoroughly discussed in the Ref.[\onlinecite{pseudo1}], the strain tensor is predominantly governed by out-of-plane displacements, given by  $\eta(x)=z^\prime(x)^2/2$. By setting the height profile to $z(x)=h_0 \exp(-x^2/2\lambda^2)$, the resulting pseudo-magnetic field is calculated via\cite{Ando,pseudo1},}
\begin{equation}
B_{\rm eff}(x)=\frac{\hbar\beta \eta^\prime(x)}{2ae}\hat{\mathbf{k}},
\label{bfield}
\end{equation}
\comment{where $\beta=3$. The pseudo-magnetic field as a function of position is shown in Supplementary Fig. \ref{magnet}, where we set $h_0=50a$ and $\lambda=60a$. The varying effective pseudomagnetic field is as large as 300 T. However, we must emphasize that this is an overestimation, as the strain tensor $\eta(x)$ reaches 0.2. Such a large strain tensor is unlikely to appear in real samples where the out-of-plane profile varies more gradually. In the following, we show that such pseudomagnetic fields can indeed alter the band structure of gapped bilayer graphene, but they cannot account for the emergence of topological channels observed inside the gap.}
\begin{figure}[H]
\centering
\includegraphics[width =0.4\linewidth]{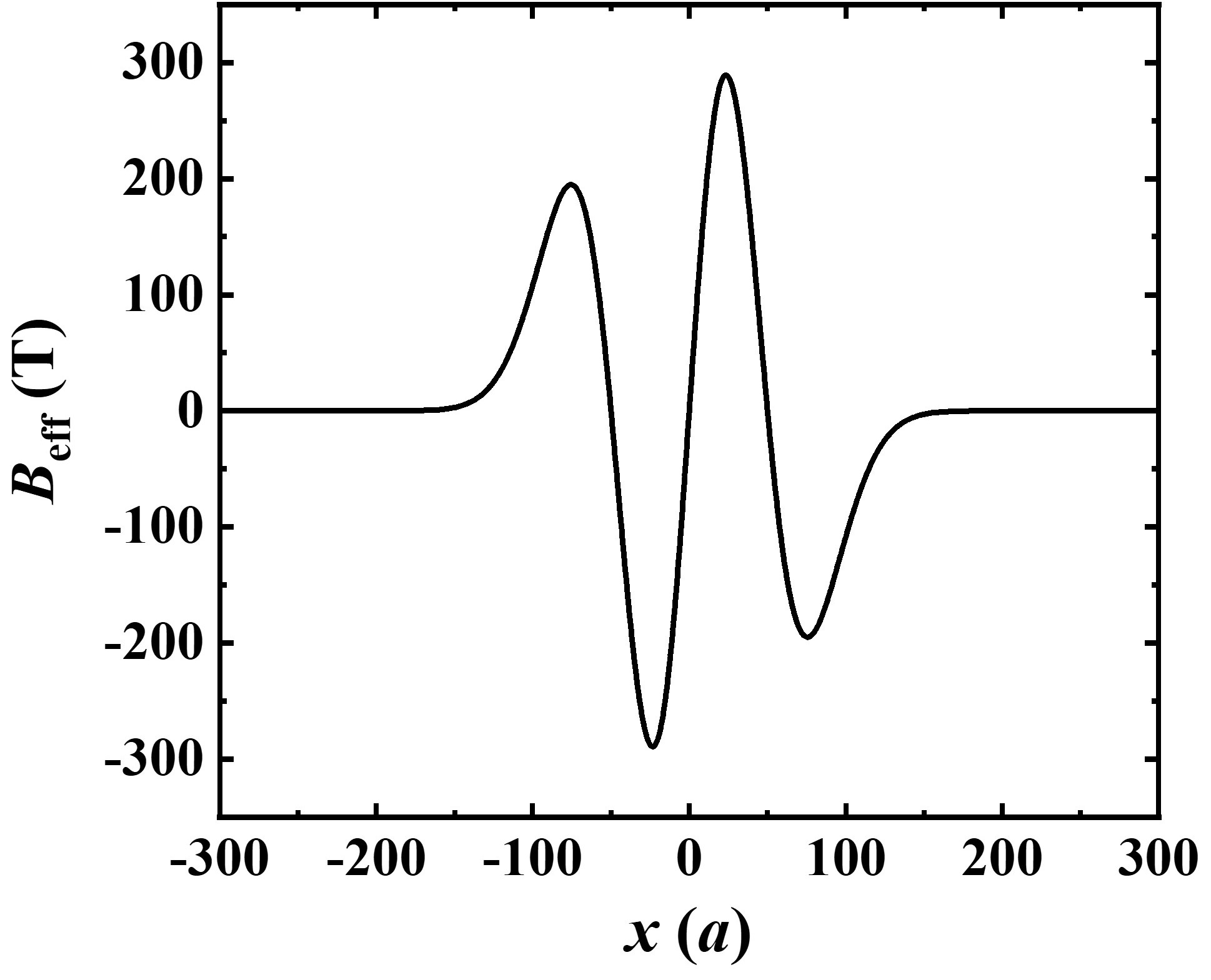}
\caption{\textbf{Pseudo-magnetic field as a function of position.}}
\label{magnet}
\end{figure}
\comment{We performed tight-binding simulations incorporating bond length variations between carbon atoms induced by out-of-plane deformations, while neglecting the effect of interlayer sliding. In these simulations, we consider a ribbon of width $800a$ and apply an interlayer potential difference of $V_\pm = \pm 0.1t$. The band structure of bilayer graphene without pseudomagnetic field is summarized in Supplementary Fig. \ref{banddeform}(a). When the pseudomagnetic field is included, the band structure undergoes a significant modification, exhibiting reduced dispersion and flatter bands, as illustrated in Supplementary Fig.\ref{banddeform}(b). This band flattening resembles Landau-level-like features, possibly induced by the inhomogeneous pseudomagnetic field, despite its spatial nonuniformity.}

\begin{figure}[H]
\centering
\includegraphics[width =0.7\linewidth]{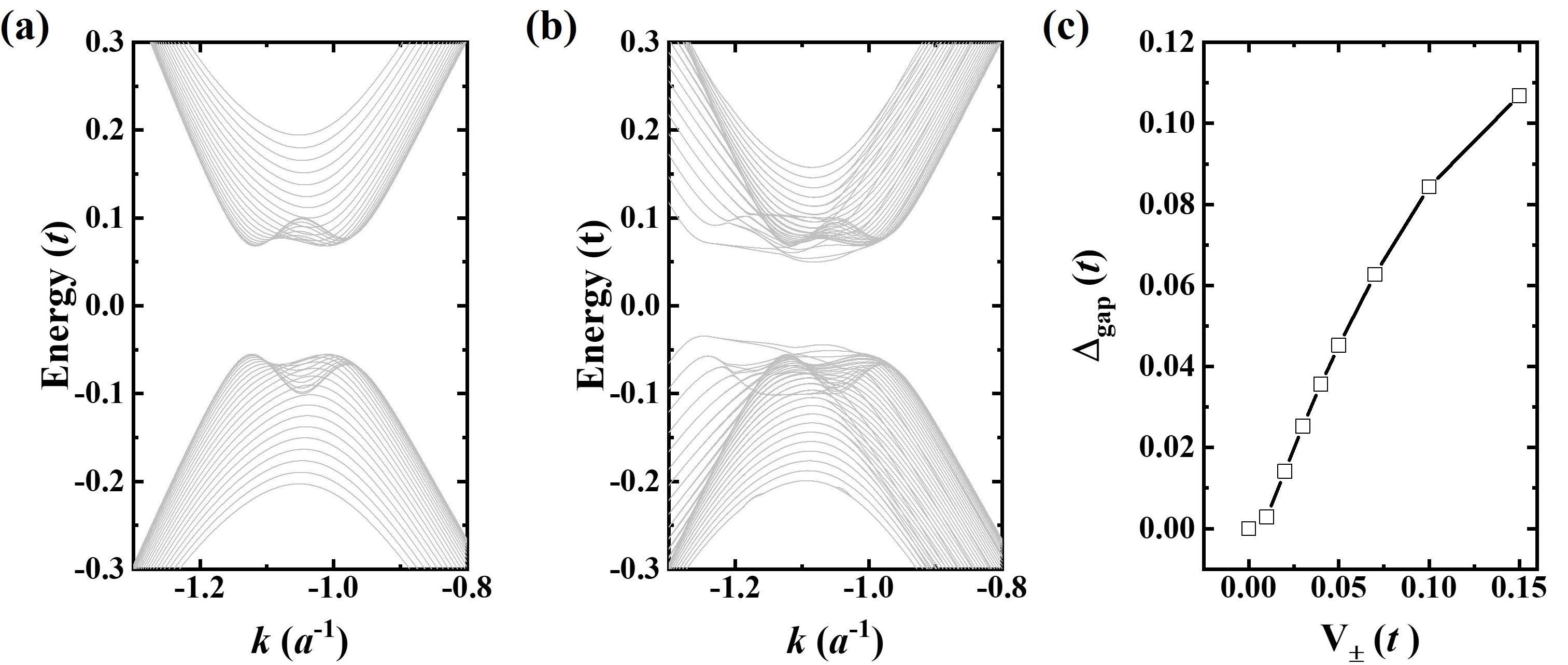}
\caption{\textbf{Bandstructure of bilayer graphene (a) without pseudomagnetic field (b) with pseudomagnetic field. (c) Summarized band gap with pseudomagnetic field for different layer potential differences.}}
\label{banddeform}
\end{figure}
\comment{Another notable effect of the pseudomagnetic field is a reduction of the band gap, as seen by comparing Supplementary Figs.\ref{banddeform}(a) and \ref{banddeform}(b). However, even at small interlayer potential differences, the pseudomagnetic field alone is insufficient to fully close the gap. We further analyze the band gap as a function of $V_\pm$, and find that the gap increases monotonically with the interlayer potential difference, as shown in Supplementary Fig. \ref{banddeform}(c). This behavior would lead to an increase in resistance with increasing $V_\pm$, which contradicts our experimental observations. We therefore conclude that the pseudomagnetic field may partially explain why our measured gap of bilayer graphene is smaller than theoretical predictions but cannot account for the emergence of the topological states observed in our experiments.}
\clearpage
\newpage
\section{Additional experimental measurement results} 
In this section, we present additional experimental measurement results, including Raman measurements to determine the lattice orientations, other source-drain and 2-probe measurement results.

\subsection{Raman measurements}
To determine the orientation of the carbon lattice, we select a bilayer graphene flake with edges forming angles of 30, 90, and 150 degrees. Following the methodology outlined in Ref. [\onlinecite{Raman1}], we scan a 2D map of the Raman D band, with the polarization direction aligned along the angle bisector, as illustrated in Supplementary Fig. \ref{raman}. The side exhibiting significantly stronger D band signals is identified as the armchair direction, while the other side corresponds to the zigzag direction.

\begin{figure}[H]
\centering
\includegraphics[width =0.5\linewidth]{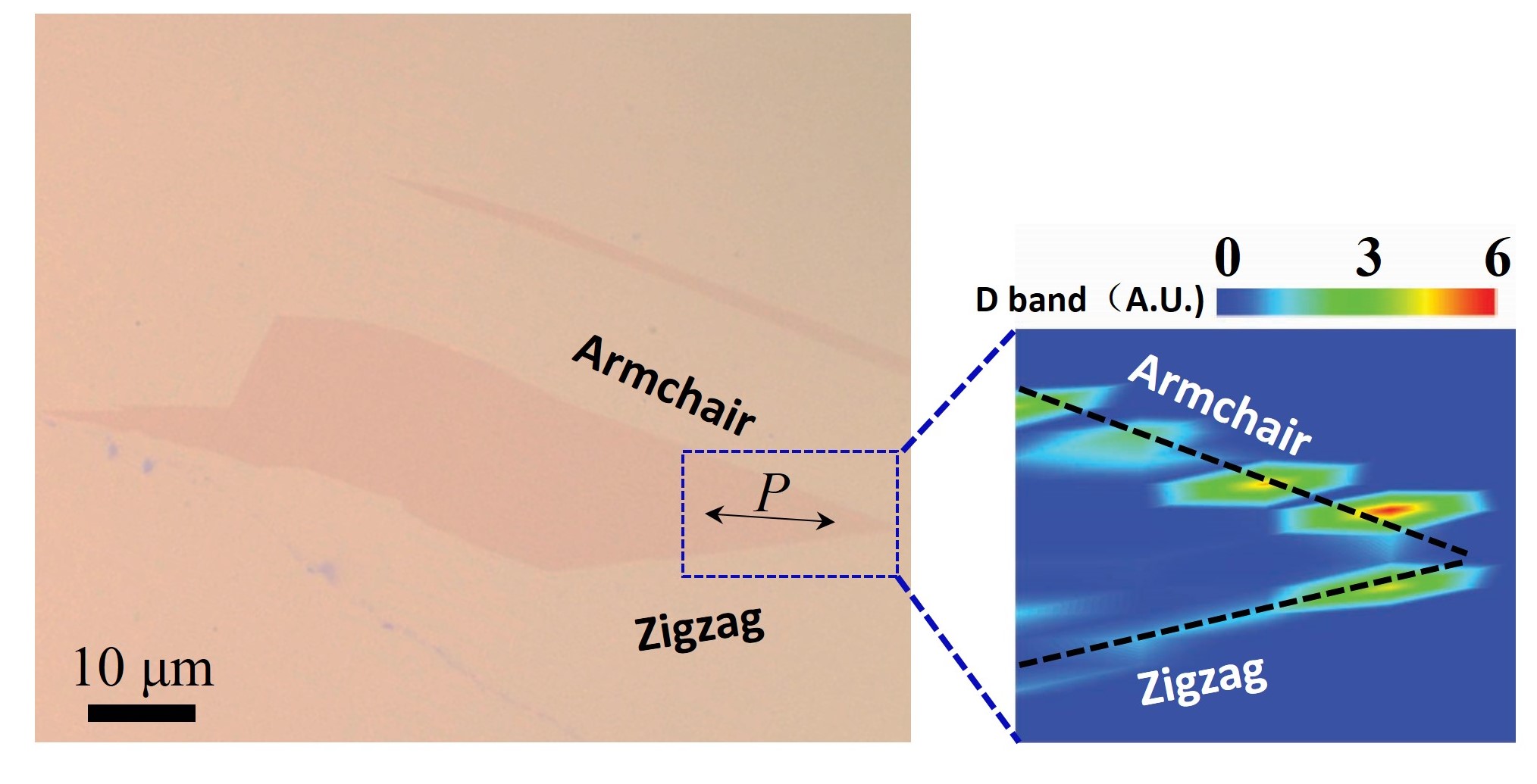}
\caption{\textbf{Raman measurements to determine the carbon lattice orientations of bilayer graphene.}}
\label{raman}
\end{figure}

\subsection{Additional source-drain measurements}
\begin{figure}[H]
\centering
\includegraphics[width =0.9\linewidth]{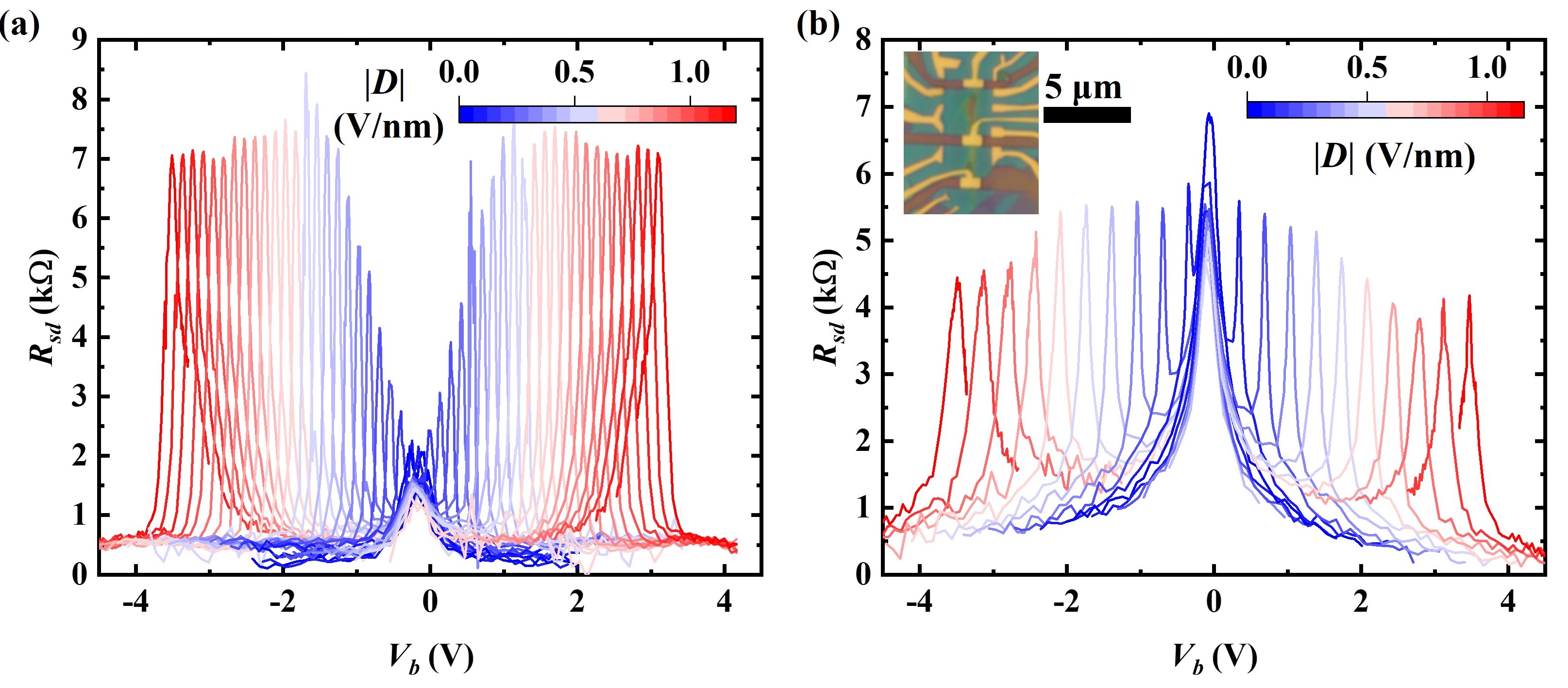}
\caption{\textbf{Source-drain measurements of Sample 1.}The source-drain resistance is plotted as a function of the back gate voltage, with color indicating the strength of displacement field. The resistance at CNP is observed to saturate around (a) 7.1 k$\Omega$ with channel length 1 $\mu$m and (b) 4.4 k$\Omega$ with channel length 0.8 $\mu$m. The third region's data is shown in  Fig. 4 in the main text.}
\label{s1}
\end{figure}

Here, we present additional source-drain measurement results for Samples 1 to 5, as shown in the following Supplementary Figs. \ref{s1}-\ref{s5}. All show saturating behavior at large displacement fields.

\begin{figure}[H]
\centering
\includegraphics[width =0.5\linewidth]{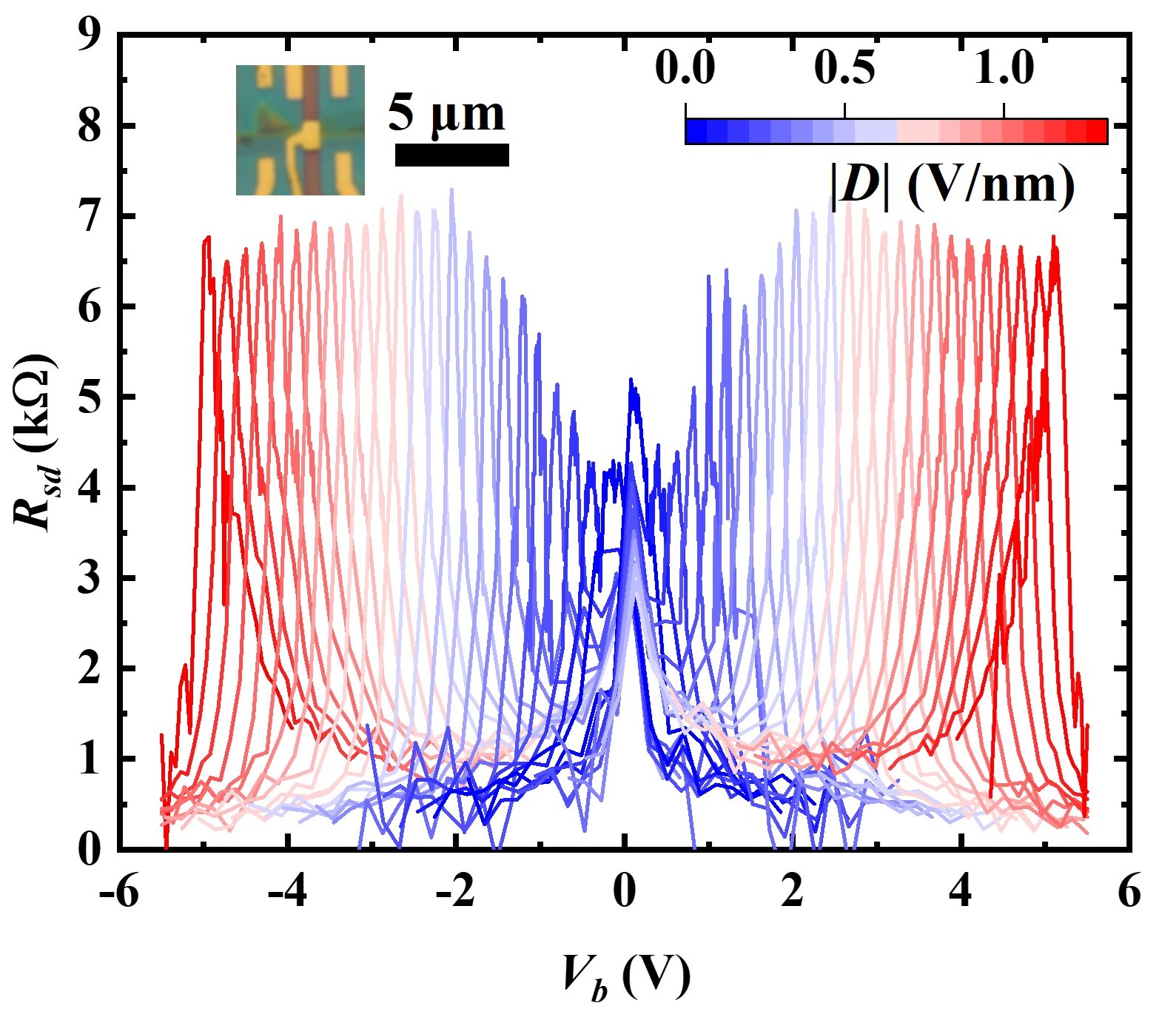}
\caption{\textbf{Source-drain measurements of Sample 2.}The source-drain resistance is plotted as a function of the back gate voltage, with color indicating the strength of displacement field. The resistance at the CNP is observed to saturate around 6.8 k$\Omega$  with channel length 0.8 $\mu$m.}
\end{figure}

\begin{figure}[H]
\centering
\includegraphics[width =0.5\linewidth]{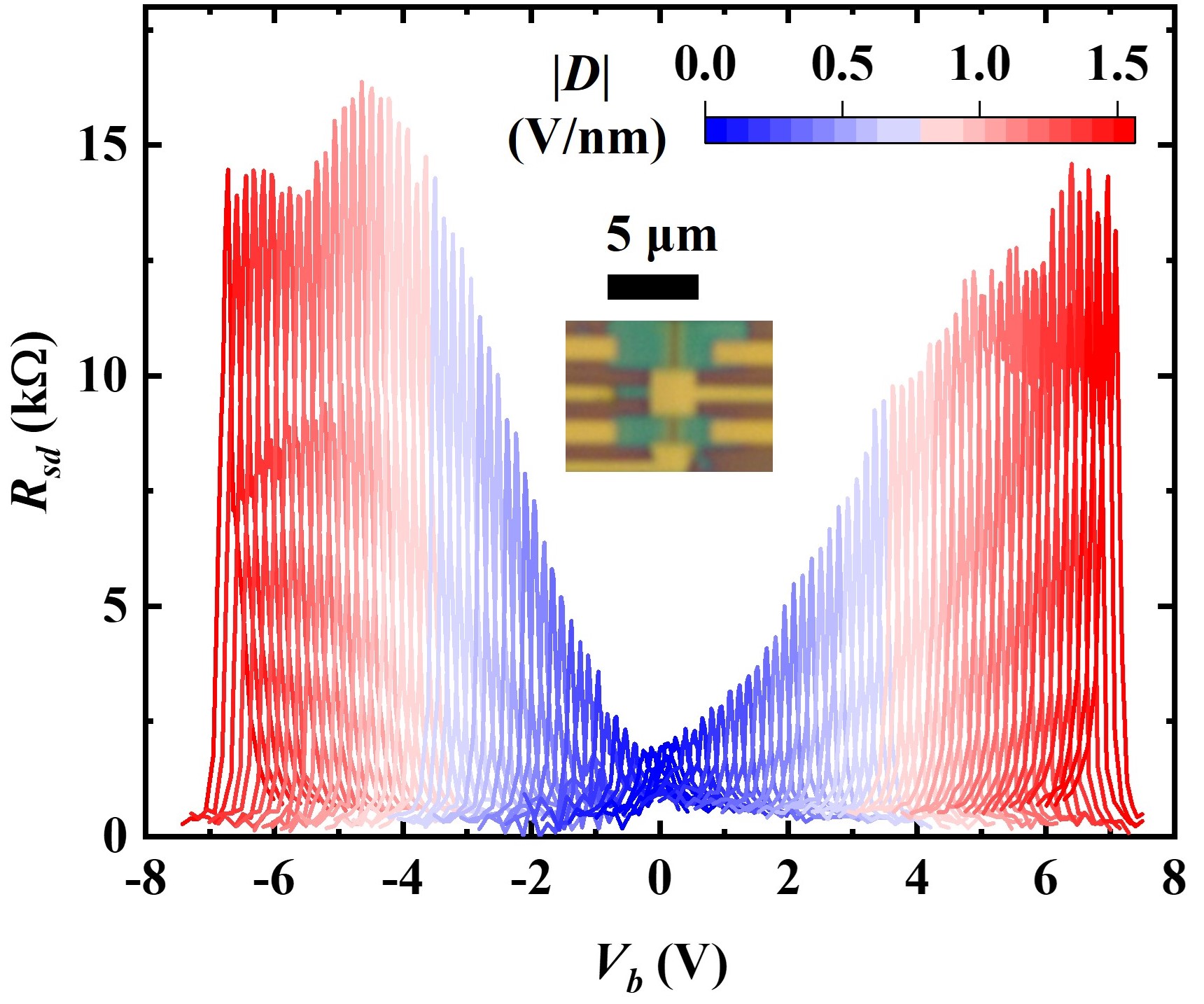}
\caption{\textbf{Source-drain measurements of Sample 3.}The source-drain resistance is plotted as a function of the back gate voltage, with color indicating the strength of displacement field. The resistance at the CNP is observed to saturate around 14 k$\Omega$ with channel length 2.5 $\mu$m.}
\end{figure}

\begin{figure}[H]
\centering
\includegraphics[width =0.5\linewidth]{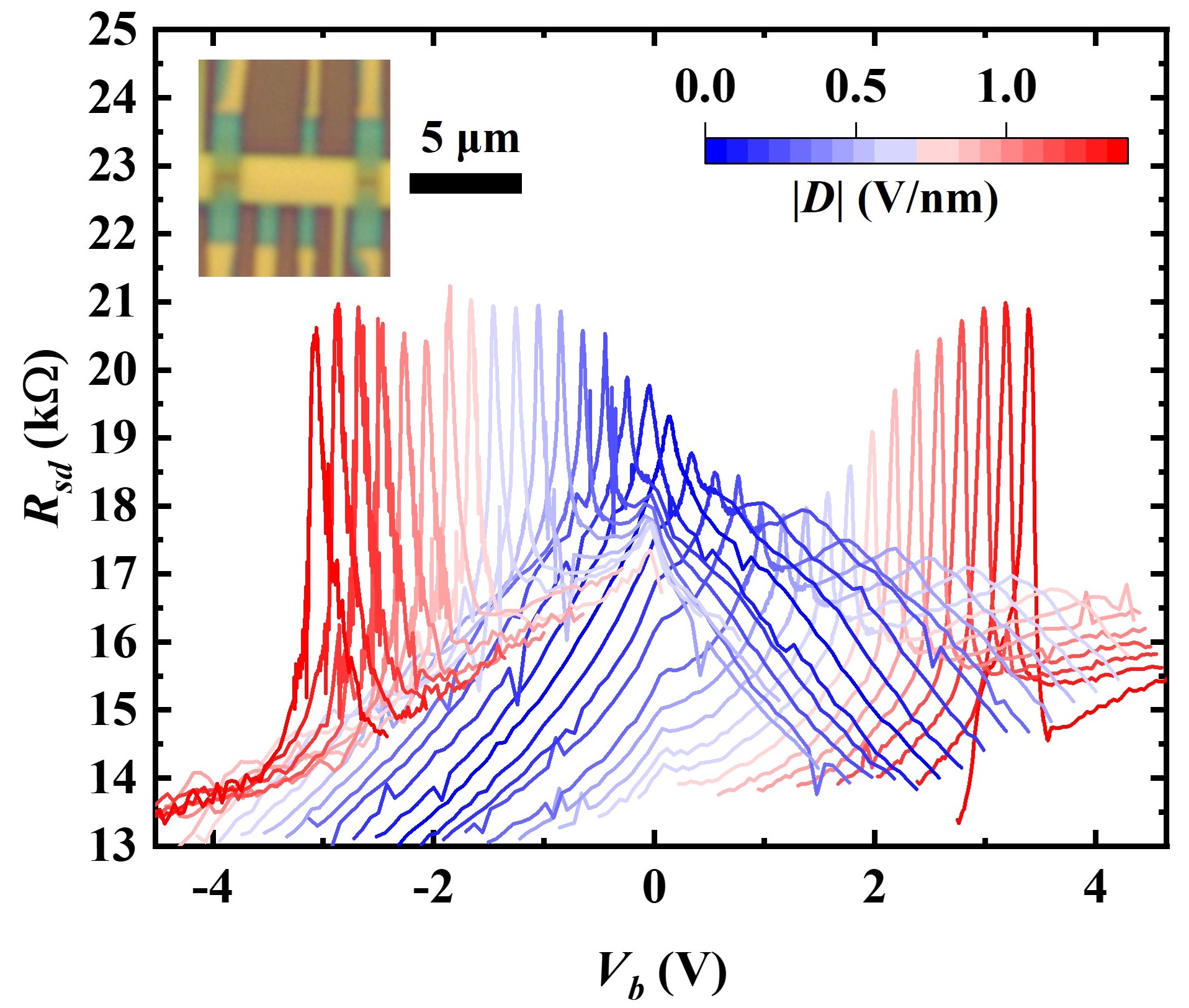}
\caption{\textbf{Source-drain measurements of Sample 4.} The source-drain resistance is plotted as a function of the back gate voltage, with color indicating the strength of displacement field. The resistance at the charge neutrality point (CNP) is observed to saturate around 21 k$\Omega$ with channel length 5 $\mu$m.}
\end{figure}

\begin{figure}[H]
\centering
\includegraphics[width =0.5\linewidth]{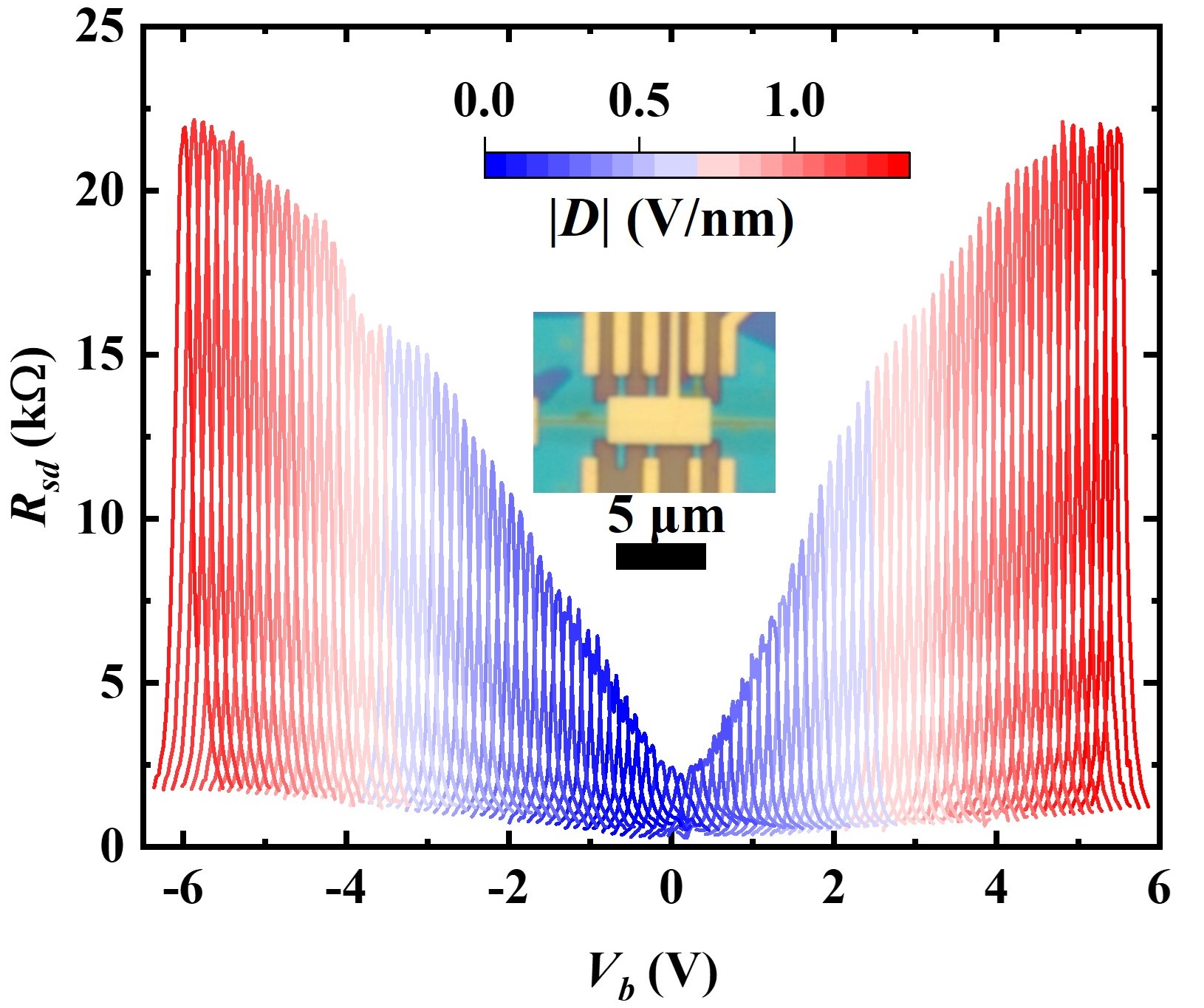}
\caption{\textbf{Source-drain measurements of Sample 5.}The source-drain resistance is plotted as a function of the back gate voltage, with color indicating the strength of displacement field. The resistance at the CNP is observed to saturate around 22 k$\Omega$ with channel length 6 $\mu$m.}
\label{s5}
\end{figure}

\comment{For the above source-drain resistance measurements, we employed a four-probe configuration to eliminate the “conventional contact resistance” between the Cr/Au electrodes and graphene, as shown in Supplementary Fig. \ref{contactR}(a) and the inset of Fig. 4 in main text. The remaining contact resistance, which is not intrinsic to the topologically protected conducting channels, is attributed to straying current effects in regions not covered by the top gates (illustrated by the dashed line in Supplementary Fig. \ref{contactR}(a)). Our finite element simulations using COMSOL (Supplementary Fig. \ref{contactR}(b)) demonstrate that when the current $I$ is injected between the source and drain (corresponding to the topological channels in the experimental devices) electrodes, the remaining contact resistance can be defined as $(V_{\rm probe} - V_{\rm drain})/I= \alpha R_\Box$. Here, $R_\Box$ stands for the sheet resistance of bilayer graphene with square geometry. The coefficient $\alpha$ is geometry-dependent and ranges from 0.5 to 0.9 for our samples. Since our analysis focuses on regions with large displacement fields where the carrier density exceeds $5\times10^{12}/{\rm cm}^2$, $R_\Box$ remains below 100 $\Omega$. This low sheet resistance has allowed us to justifiably neglect the contact resistance in our Landauer-Büttiker formalism analysis. We notice that contact resistance is also neglected in the similar way in previous reports\cite{Ju2015,Li2016,Campos2019}}.

\begin{figure}[H]
\centering
\includegraphics[width =0.5\linewidth]{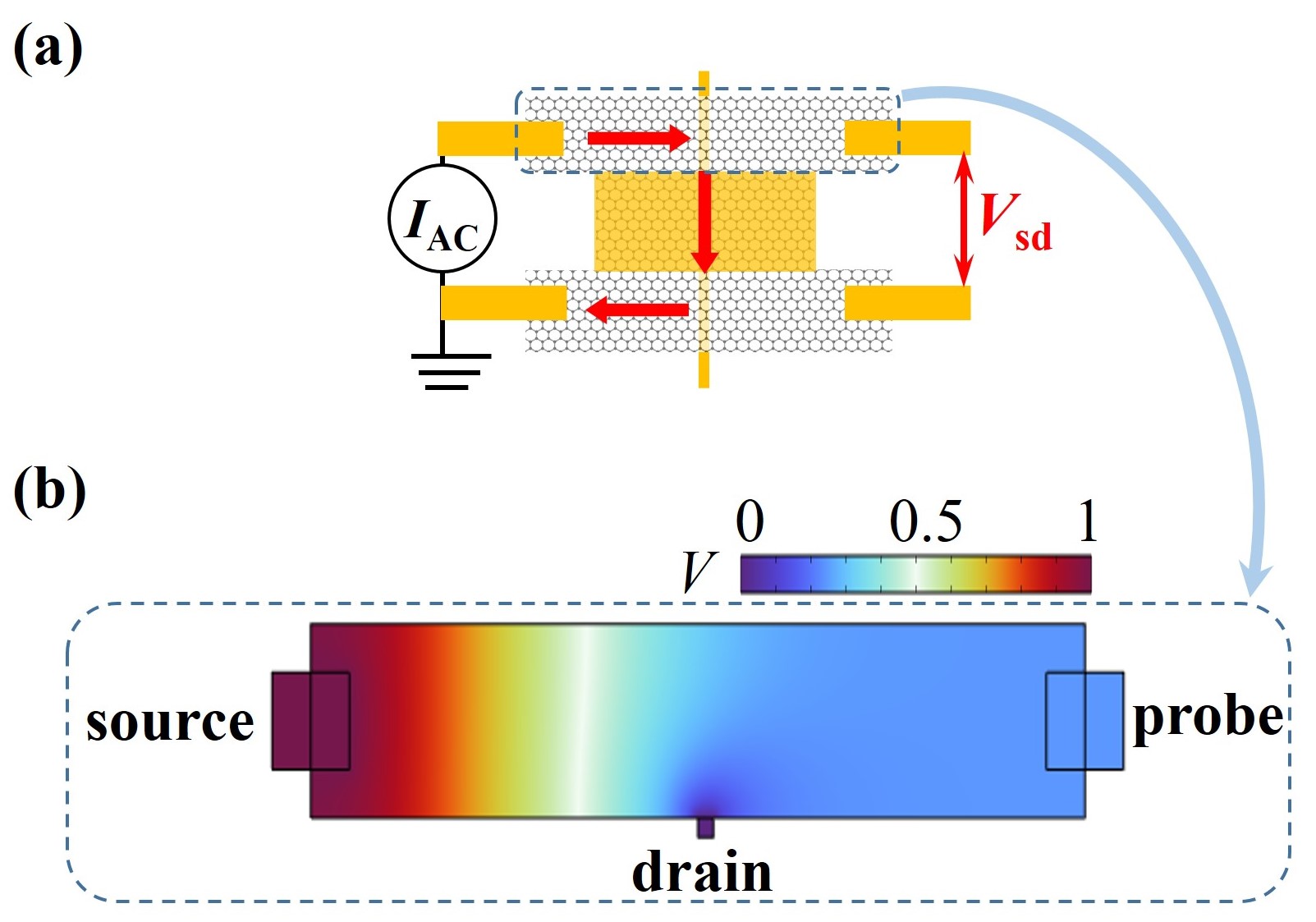}
\caption{\textbf{Source-drain measurements of Sample 5.}The source-drain resistance is plotted as a function of the back gate voltage, with color indicating the strength of displacement field. The resistance at the CNP is observed to saturate around 22 k$\Omega$ with channel length 6 $\mu$m.}
\label{contactR}
\end{figure}

Approximately 70\% of fabricated samples display a saturating resistance at high displacement fields. In contrast, the remaining 30\% exhibit high, non-saturating resistance under large displacement fields. One representative non-saturating result is shown in Supplementary Fig. \ref{no-sat}.
\begin{figure}[H]
\centering
\includegraphics[width =0.5\linewidth]{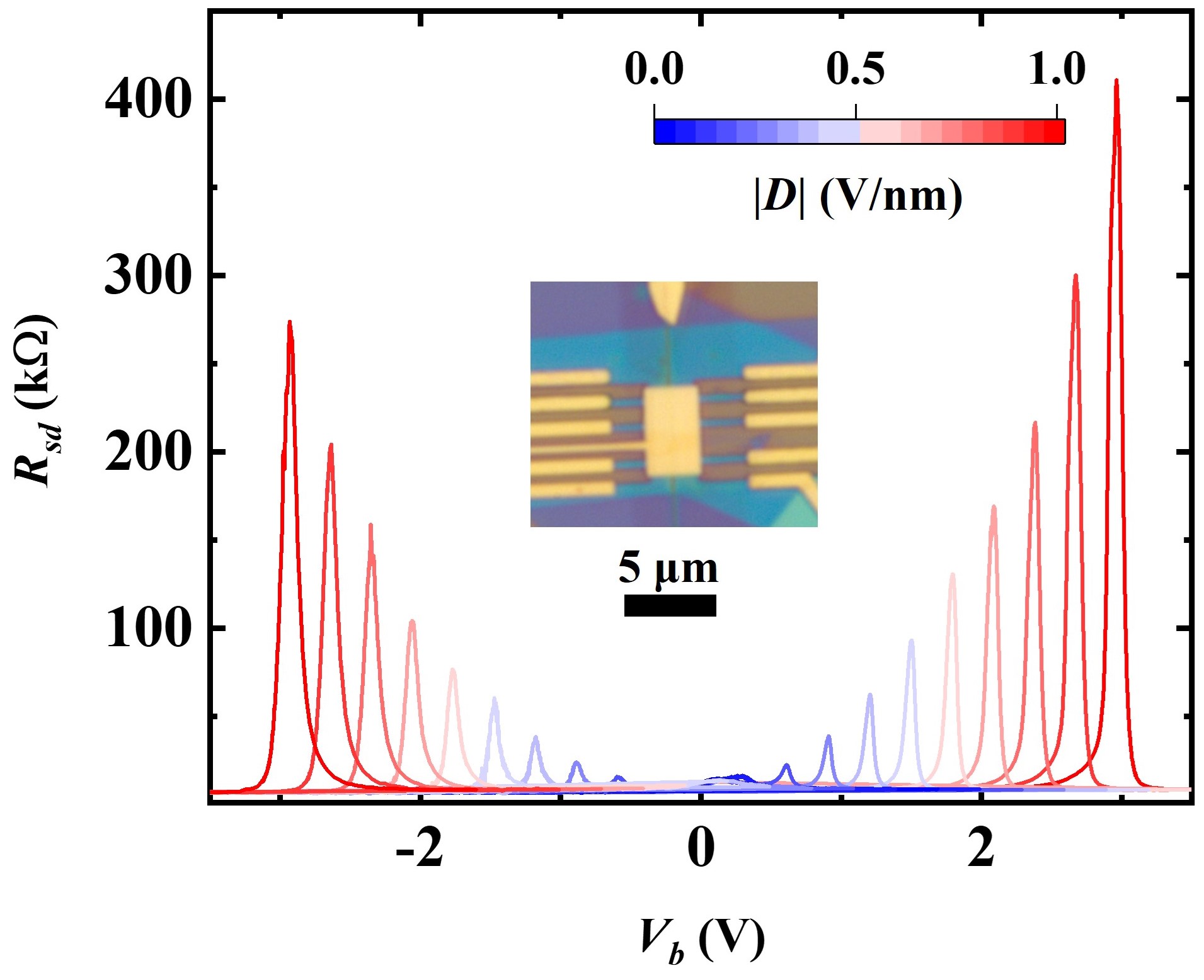}
\caption{\textbf{Source-drain measurements of Sample 6.}The source-drain resistance is plotted as a function of the back gate voltage, with color indicating the strength of displacement field. The resistance at the CNP is observed to show no saturating behavior with channel length 5 $\mu$m.}
\label{no-sat}
\end{figure}

\subsection{Excluding potential edge state contributions} 
Apart from the interlayer sliding effect discussed in the main text, the edges of bilayer graphene can support topological states\cite{Zhu2017,subgap2022}. However, these topological states can be scattered by random atomic defects at the sample edges, resulting in a relatively high resistance, around hundreds of k$\Omega$ for $L$ at sub 10 $\mu$m scale\cite{Ju2015,Zhu2017}. Typically, to observe the topological transport behavior of sample edge states, the channel length needs to be much shorter than in our case, usually less than 100 nm, as reported in Ref. [\onlinecite{subgap2022}]. Furthermore, these edge states can support four chiral channels from two edges, leading to a minimum CNP resistance of $h/4e^2 = 6.4$ k$\Omega$, which is larger than the observed resistance of 4.7 k$\Omega$. To further exclude the influence of edge state mechanisms, we perform 2-probe measurements using different sets of electrodes of Sample 5. One measurement type involves the 2-probe resistance between the source (``1'') and drain (``7") electrodes, labeled ``1-7" in Supplementary Fig. \ref{s5-1}(a). The other 2-probe resistance measurements at the CNP are conducted between the Hall electrodes, labeled ``2-3", ``3-4", and ``5-6". Notably, the resistance between the source and drain electrodes (``1-7") exhibits the longest channel length yet shows the smallest 2-probe resistance compared to the Hall probes. More importantly, as the displacement field $D$ increases,  the 2-probe resistance ``1-7" exhibits a saturation behavior, which is absent for the resistance measured between the Hall pairs. In contrast to the symmetric 2-probe source-drain resistance, the resistance measured between the Hall electrodes displays asymmetric behavior, resembling the transport characteristics typically associated with topological edge states, as detailed in Ref. [\onlinecite{subgap2022}]. 

\begin{figure}[H]
\centering
\includegraphics[width =0.8\linewidth]{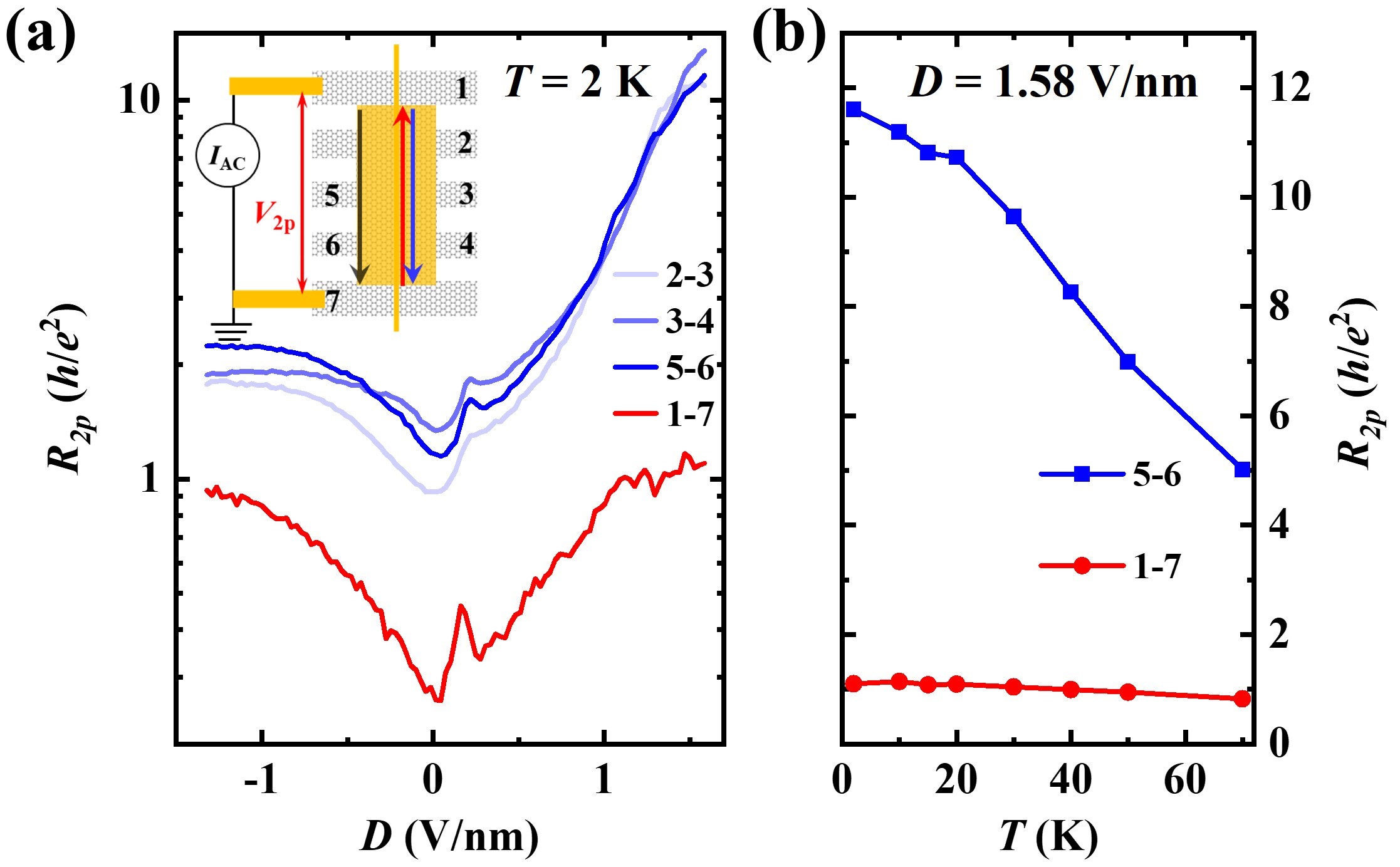}
\caption{\textbf{2-probe measurement of Sample 5.} (a) CNP resistance as a function of displacement field for different electrode pairs. All the measurements are conducted at 2 K. (b) CNP resistance as a function of temperature for fixed displacement field $D=1.58$ V/nm.}
\label{s5-1}
\end{figure}

In addition, we also investigate the temperature dependence of the CNP resistance as summarized in Supplementary Fig. \ref{s5-1}(b). The source-drain resistance remains almost unchanged when the temperature cools down to 2 K, while the resistance between hall electrode pairs "5-6" increases dramatically. These different characteristics, as shown in Supplementary Fig. \ref{s5-1}, collectively suggest that the edge states at the sample edges cannot account for the observed topological transport.

\clearpage
\newpage

\bibliography{biblio}